\RequirePackage{fix-cm}
\documentclass[smallextended]{svjour3}       
\smartqed  
\usepackage{graphicx}
\usepackage{mathrsfs}
\usepackage{bm}

\usepackage{ntheorem}
\theoremseparator{:}

\usepackage{amsmath}

\usepackage[caption=false]{subfig}

\usepackage[dont-mess-around]{fnpct}

\usepackage{etoolbox}
\newcommand*{\affaddr}[1]{#1} 
\newcommand*{\affmark}[1][*]{\textsuperscript{#1}}

\usepackage{lineno}

\renewcommand{\Psi}{{\varPsi}}
\def\diag{\mathop{\mathrm{diag}}}

\def\Xint#1{\mathchoice
      {\XXint\displaystyle\textstyle{#1}}%
      {\XXint\textstyle\scriptstyle{#1}}%
      {\XXint\scriptstyle\scriptscriptstyle{#1}}%
      {\XXint\scriptscriptstyle\scriptscriptstyle{#1}}%
      \!\int}
   \def\XXint#1#2#3{{\setbox0=\hbox{$#1{#2#3}{\int}$}
        \vcenter{\hbox{$#2#3$}}\kern-.5\wd0}}
   
   \def\dashint{\Xint-}

\begin{document}

\title{Statistical Measures and Selective Decay Principle for Generalized Euler Dynamics: 
       Formulation and Application to the Formation of Strong Fronts
}

\titlerunning{Generalized Euler Dynamics}        

\author{Giovanni Conti\protect\affmark[1] \protect\affmark[2]    \and
        Gualtiero Badin\protect\affmark[1] 
}


\institute{Giovanni Conti \at
              \email{giovanni.conti@uni-hamburg.de}         
           \and
           Gualtiero Badin \at
           \email{gualtiero.badin@uni-hamburg.de} \\ \\   
\affaddr{\affmark[1]Center for Earth System Research and Sustainability (CEN), University of Hamburg, Hamburg, Germany}\\
\affaddr{\affmark[2]Foundation Euro-Mediterranean Center on Climate Change (CMCC), Bologna, Italy}           
}

\date{Received: date / Accepted: date}

\maketitle

\begin{abstract}

In this work we investigate the statistical mechanics of a family of two dimensional (2D) fluid flows, described by the generalized Euler equations, or $\alpha$-models. These models describe both nonlocal and local dynamics, with one example of the latter given by the Surface Quasi Geostrophy (SQG) model for which the existence of singularities is still under discussion. 
Furthermore, SQG is relevant both for atmosphere and ocean dynamics, and in particular, it is proposed to understand the oceanic submesoscale structures, front and filaments, associated with horizontal gradients of buoyancy.
We discuss under which conditions the statistical theory suggests a principle of selective decay for the whole family of turbulent models, and then we  explore the selective decay principle numerically, the transition to equilibrium and the formation of singularities, starting from initial conditions (i.c.s) corresponding to a hyperbolic saddle. 
 We study the   topological transitions in the flow configuration induced by filaments breaking. Furthermore we compare the  theoretical equilibrium states, and the functional relation between the generalized vorticity $q$ and its correspondent stream function $\psi$, with the results of the simulations. For the particular i.c.s investigated, and domain used for the simulations, we have not  noticed  transition from tanh-like to sinh-like $\psi-q$ functional relation, which would be expected by the emergence of coherent structures that are, however, filtered by the time averages used to compute the mean fields.

\end{abstract}

\section{\label{sec:level1} Introduction} 

Despite its importance, turbulence still remains one of the unsolved problems of physics. However, a statistical approach can be useful and profitable in the characterization of fluids undergoing turbulent motion \cite{Kraichnan1967,kraichnan_1975,frisch_1995}. 

In this study we investigate the equilibrium theory of a family of two dimensional ($2$D) incompressible fluid models that  exhibit turbulent behavior by using a statistical mechanics approach and their relation to the possible formation of singularities.
These sources of information allow us to introduce a probability measure and a mean field equation that functionally  relate the stream function and the active scalar, in other words, they provide a constraint under which the solutions undergoing turbulent motion will have to tend, and relax, towards equilibrium. 

The governing equation for the family of models is
\begin{equation}
\frac{\partial q}{\partial t} +\mathcal{J}(\psi,q)=0
\label{eq1}
\end{equation}
on the plane $\mathcal{R}^2$. In \eqref{eq1},  $q(x,y,t)$ is an active scalar,  or generalized potential vorticity, $J(\psi,q)=\partial_x \psi \partial_y q - \partial_x q \partial_y \psi$ is the Jacobian determinant, and $\psi(x,y,t)$ is the stream function which fulfills the relationship 
\begin{equation}
q=-(-\Delta)^{\alpha/2}\psi,
\label{invp}
\end{equation}
with the parameter $\alpha\in \mathcal{R}$ that determines the degree of locality, or the range of interaction, of the model. These equations are known as generalized Euler equations or  $\alpha$-models (that must not be confused with Euler-$\alpha$ models \cite{HOLM19981,badin2018geometric}).
The effect of the $\alpha$-parameter can be understood better  transforming \eqref{invp} in the Fourier space,
\begin{equation}
\hat{\psi}(\vec{k})=-\left\lVert\vec{k}\right\rVert^{-\alpha}\hat{q},
\label{invf}
\end{equation}
where $\left\lVert \cdot\right\rVert$ is the usual $L^2$ norm, and $\vec{k}$ is the $2$D wavenumber vector. When $\alpha$ increases, fields become more decoupled.  For $\alpha < 2$ the dynamics is said to be local, while for $\alpha > 2$ the dynamics is said to be non-local. These models were introduced by  \cite{pierrehumbertetal94} and have  been studied for several values of $\alpha$ by e.g. \cite{weinstein1989time,smith2002turbulent,tran2002constraints,tran2004nonlinear,bernard2007inverse,tran2010effective,burgess2013spectral,burgess2015kraichnan,Norbert2000,Venaille2015,foussard2017relative,BadinBarry2018,Conti2019}.

In this work we consider the cases $\alpha=1$ and $\alpha=2$.
When $\alpha=2$, \eqref{eq1} and \eqref{invp} describe the widely studied Euler equations with the active scalar representing the vorticity. 
When $\alpha=1$  \eqref{eq1} and \eqref{invp} can be used to study the local  dynamics of a stratified rapidly rotating flow upon a surface bounding a constant interior of potential vorticity   (e.g. the atmospheric tropopause or the oceanic surface). In this case the scalar field $q$ takes the meaning of potential temperature or buoyancy. This model is called Surface Quasi Geostrophic (SQG) model \cite{blumen78,heldetal95,lapeyre17,badin2018variational}. 
The SQG approximation, in the atmospheric dynamics context, has been considered to study the dynamical properties of potential temperature anomalies on the tropopause \cite{Juckes1994} and the asymmetry between cyclones and anticyclones \cite{Hakim2002}. The shape of the tropopause spectra can also be explained by means of this approximation \cite{tulloch2006theory}.
In the ocean dynamics context SQG has been used to infer the interior dynamics given the sea surface temperature  anomalies \cite{LaCasce2006,Lapeyre2006,Isern2008,LaCasce2012,Lapeyre2009}.
SQG exhibits  a characteristic forward surface kinetic energy cascade \cite{tulloch2006theory,capet2008surface}, with consequent formation of frontal structures in the flow.  
For these reasons, SQG is also a candidate for the explanation of the submesoscale dynamics (where the term submesoscale is here used in its oceanographic meaning, see e.g. \cite{McWilliams2016}). 
These small scale structures, fronts and filaments, are important for the high vertical velocities associated with them, for their large values of relative vorticity and for their strong contribution to vertical fluxes of properties \cite{Klein2008,Capet2008}.
As a consequence of this, SQG is also  considered important for the mixing of passive tracers in the atmosphere and in the ocean \cite{badin2011lateral,Shcherbina2015,Mukiibi2016}.  The role of principles related to this kind of dynamics could be important in order to understand the climate system \cite{Levy2010}.
 For a study on the relationship between quasi geostrophic (QG) and SQG turbulence, see e.g. \cite{badin2014role}. For the Hamiltonian and Nambu structure of SQG, see e.g. \cite{blender2015hydrodynamic,badin2018variational}.

From a mathematical point of view, SQG shows interesting analogy with the $3$D Euler equation \cite{constantinetal94}. This analogy  suggests that the study of the regularity of the SQG model could provide hints for the formation of singularities in the 3D Euler equation \cite{pierrehumbertetal94,constantinetal94,constantin1994singular,majda1996two,ohkitani1997inviscid,cordoba1998nonexistence,constantin1998nonsingular,constantin1999behavior,cordoba2002growth,cordoba2002scalars,cordoba2004maximum,rodrigo2004vortex,cordoba2005evidence,rodrigo2005evolution,wu2005solutions,deng2006level,dong2008finite,ju2006geometric,li2009existence,marchand2008existence,marchand2008weak,scott2011scenario,constantin2012new,ohkitani2012asymptotics,scott2014numerical,BadinBarry2018,scott2019scale}.

Although these models have the tendency to generate turbulence,  they also have the ability to reorganize the  fine-scale turbulence in large-scale coherent structures thus in a sort of macroscopic order. 

Regarding $\alpha = 2$, different statistical approaches have been considered for the explanation of the formation of these coherent structures.
Bretherton and Haidvogel \cite{bretherton_haidvogel_1976} proposed a phenomenological approach, later called selective decay hypothesis  \cite{Matthaeus1980}, to explain the equilibrium state of a fluid. Their hypothesis  is that, at the equilibrium, the flow will tend to minimize the enstrophy of the  system. A more formal justification can be found in \cite{majda_wang_2006}.
It is legitimate to ask whether this type of hypothesis also applies to the system of generalized Euler equations.

One of the first attempts to use proper statistical mechanics of 2D turbulence was proposed by Onsager \cite{Onsager1949}, and the theory was further developed by e.g. \cite{joyce_montgomery_1973,Lundgren1977,caglioti1992,caglioti1995}. This approach is based on the approximation of the continuous system by means of point vortices. This leads to the study of a Hamiltonian system for which the Liouville theorem holds, and it is therefore possible to employ statistical mechanics. However this is a drastic simplification of the original system.

Another mechanical statistical approach was introduced by Kraichnan \cite{Kraichnan1967,kraichnan_1975}. This method is based on the approximation of the continuous system by means of a truncation in the Fourier series representing the vorticity field. In this case the resulting system respects the Liouville theorem and it is therefore possible to develop a proper statistical mechanics. This method also shows some problems. In fact truncation leads to the loss of system invariants. This theory is called Energy-Enstrophy statistical theory, since energy and enstrophy are the only conserved quantities that are used to construct the theory. Attempts to extend the method with limit procedures, considering infinite number of components of the Fourier series \cite{Boldrighini1980}, have produced results which are valid only at the formal level.

A theory that considers the continuous vorticity field has been developed by Miller \cite{Miller1990}, Robert and Sommeria \cite{robert_sommeria_1991} (MRS). In this theory, that exploits all the motion invariants, the organized structures appear as states of maximum entropy. 
It takes into account a  more general point of view of statistical mechanics. 
 The functional relations between $q$ and $\psi$ are influenced by the initial conditions through the Lagrange multipliers used in the maximization process and related to the motion invariants.
 Miller \cite{Miller1990} showed how from particular initial conditions, which lead to a mean Gaussian vorticity distribution at the statistical equilibrium, a linear relationship emerges between vorticity and stream function. Chavanis and Sommeria \cite{chavanis_sommeria_1996} showed how in the limit of strong mixing,  the maximization of the entropy becomes equivalent to the minimization of the enstrophy,  and they established a hierarchy between the constraints of the system, showing that only  energy and circulation are  important in those situations.

Another theory valid for continuous system was introduced by \cite{Ellis_2002} and further discussed by \cite{Chavanis2005,Chavanis2008,Chavanis2010}.  Since in real situations some of the constraints are not conserved due to the presence of forcing and/or dissipation, in the modified MRS theory some of the constraints and biases are taken into account by means of a prior distribution. This theory provides a way to define an equilibrium measure and a functional relation between the active scalar and the stream function. Furthermore  it easily suggests and formalizes a selective decay principle for the entire family of $\alpha$-models. That is, it provides a hint regards the relaxation of the system toward the equilibrium. 


In this work, while taking advantage of the continuum theory that generalizes the MRS theory,  we also show how the principle of selective decay simply emerges for all values of $\alpha$. Although this result is based on a particular prior distribution, this principle seems to have a more general value as shown by using numerical simulations.
 The relaxation toward the equilibrium for different $\alpha$-models was previously explored by \cite{Venaille2015}, but with the use of different initial conditions i.c.s. Our work formalizes part of the work reported by \cite{Venaille2015} and should be read as complementary to it.
We explore the selective decay associated to a hyperbolic saddle geometry, which has aroused interest since early studies have suggested that it could be able to develop singularities for the inviscid $\alpha=1$ (SQG) case \cite{constantinetal94,constantin1994singular,constantin1998nonsingular,constantin2012new}. 
The motivation, to investigate singularities by considering a method derived by statistical mechanics,  comes from the research for parallelisms between singularity, filament breakage and phase transitions.
In order to explore the space of i.c.s we also study perturbed initial states. 



The paper is organized as follows.  In Section \ref{sec:level3} we review a statistical theory for the continuous system, and we show how this theory suggests a selective decay principle for the whole family of generalized models. 
In Section \ref{sec:level4} we  numerically investigate the selective decay principle and the functional relation between the active scalar and the stream function. In Section \ref{conc} we report the final discussion.
In the appendix  \ref{sec:level2} we show a difference in the definition of the statistical measure for the point vortex approximation, which is here of minor  importance, between local dynamics and $2$D turbulence.


%
%

\section{\label{sec:level3} Phenomenological Statistical Theory for the Generalized Euler Equations}

In the following we will proceed by using the modified MRS theory initially proposed by \cite{Ellis_2002} and further developed by \cite{Chavanis2005,Chavanis2008,Chavanis2010}.
In this theory the only constraints that are employed  are the energy and circulation, while small scales fluctuations are taken into account by means of a so called prior distribution.


Considering the continuous equations \eqref{eq1}-\eqref{invp} we study the one point statistical information of the generalized potential vorticity. 
Following \cite{majda_wang_2006} we define a probability density $\rho(\vec{z},\nu)$ on $\Omega\times\mathcal{R}^1$, where $\Omega$ is the domain of our system,
so that $\rho(\vec{z},\nu)$ is the probability density to find the active scalar between $\nu$ and $\nu+{\rm{d}}\nu$ on $\vec{z}=(x,y)$ and
\begin{equation}
\rho(\vec{z},\nu)\ge0,\quad \dashint_{\Omega}\int_{\mathcal{R}^1}\rho(\vec{z},\nu)\,{\rm{d}}\nu\,{\rm{d}}\vec{z}=1, 
\label{c1}
\end{equation}
with
\begin{equation*}
\dashint_{\Omega}f=\frac{1}{\Omega}\int_{\Omega}f \, .
\end{equation*}
Together with these conditions we consider a prior distribution $\Pi_0(\vec{z},\nu)$ thus meeting the same normalizing conditions  of $\rho$.
The entropy functional that we want to maximize must take into account  the information coming both from  $\rho$ and  $\Pi_0$. 
The natural functional to consider is then the relative entropy functional
\begin{equation}
\mathcal{S}(\rho,\Pi_0)\equiv -\dashint_{\Omega}\int_{\mathcal{R}^1}\rho(\vec{z},\nu)\log\left(\frac{\rho(\vec{z},\nu)}{\Pi_0(\vec{z},\nu)}\right){\rm{d}}\nu\,{\rm{d}}\vec{z} \, .
\label{re}
\end{equation}
We want to find the function $\rho^*$ that maximizes \eqref{re} under certain constraints $\mathcal{C}$
\begin{equation}
\mathcal{S}(\rho^*,\Pi_0)=\max_{\rho\in\mathcal{C}}\mathcal{S}(\rho,\Pi_0).
\end{equation}
We consider, in particular, the normalization of the probability density function (pdf), the mean generalized energy and the mean circulation represented using the one point statistic, thus
\begin{equation}
\mathcal{C}=\mathcal{C}_0\cap\mathcal{C}_{E}\cap\mathcal{C}_{\Gamma} \, ,
\end{equation}
where
\begin{subequations}
\begin{align}
&\mathcal{C}_0=\{\rho \,\lvert\, M(\rho)=\int_{\mathcal{R}^{1}} \rho(\vec{z},\nu)\,{\rm{d}}\nu=1 \, , \quad  \forall\vec{z} \in \Omega\} \, ,\\
&\mathcal{C}_{E}=\{\rho\,\lvert\,  E(\rho)=-\dashint_{\Omega}\bar{\psi} \bar{q}\,{\rm{d}}\vec{z}=E_0\}\, , \\
&\mathcal{C}_{\Gamma}=\{\rho \,\lvert\, \Gamma(\rho)=\dashint_{\Omega}\bar{q}\,{\rm{d}}\vec{z}
=\Gamma_0\}\, .
\end{align}
\end{subequations}
The overbar represents the average  with respect to the one point statistic.
According to the Lagrange multiplier method we have
\begin{equation}
\rho^*(\vec{z},\lambda)=\frac{\exp\left[ \nu \left( \lambda_{E}\bar{\psi^*}-\lambda_\Gamma\right)\right]\Pi_0(\vec{z},\nu) }{\int_{\mathcal{R}^1}\exp\left[ \nu \left( \lambda_{E}\bar{\psi^*}-\lambda_\Gamma\right)\right]\Pi_0(\vec{z},\nu) \,{\rm{d}}\nu } \, ,
\label{epdf}
\end{equation}
where  $\lambda_{E}$, $\lambda_{\Gamma}$ are the Lagrange multipliers of the problem, respectively related to the energy and circulation constraints.
The mean field equation for $q^*$ is then written as 
\begin{equation}
\bar{q^*}=-(-\Delta)^{\alpha/2}\bar{\psi^*}(\vec{z}) =\frac{\int_{\mathcal{R}^1}\nu\exp\left[ \nu \left( \lambda_{E}\bar{\psi^*}-\lambda_\Gamma\right)\right]\Pi_0(\vec{z},\nu) }{\int_{\mathcal{R}^1}\exp\left[ \nu \left( \lambda_{E}\bar{\psi^*}-\lambda_\Gamma\right)\right]\Pi_0(\vec{z},\nu) \,{\rm{d}}\nu } \, .\label{mfe}
\end{equation}
Defining the partition function as
\begin{equation}
\mathcal{Z}(\psi,\vec{z})=\int_{\mathcal{R}^1}\exp\left[ \nu \left( \lambda_{E}\bar{\psi^*}-\lambda_\Gamma\right)\right]\Pi_0(\vec{z},\nu) \,{\rm{d}}\nu \, ,
\end{equation}
the mean field equation \eqref{mfe} can be restated as
\begin{equation}
-(-\Delta)^{\alpha/2}\bar{\psi^*}(\vec{z})=\frac{1}{\lambda_{E}}\frac{ {\rm{\partial}} }{ {\rm{\partial}}\psi }\log\mathcal{Z}(\psi,\vec{z})\Bigg\lvert_{\psi=\bar{\psi^*}} \, .
\end{equation}

In order to comprehend the theory better, we need to specify the form of the prior distribution.

\subsection{\label{sec:level3_1_2}The Gaussian Prior Distribution and the Principle of Selective Decay}
Deeper insight can be obtained modeling the small scales fluctuations of  the generalized potential vorticity with Gaussian fluctuations with a given variance and zero mean. In fact, this kind of prior distribution generates a quadratic entropy functional \cite{Bouchette2012} that naturally suggests  the extension of the  \textit{selective decay principle} \cite{majda_wang_2006} to the entire family of the $\alpha$-models of turbulence. Another consequence of this small scale parameterization is the linear relation arising between the generalized potential vorticity and the correspondent stream function.

Let us consider a prior distribution
\begin{equation}
\Pi_0(\nu)=\sqrt{\frac{1}{2\pi\alpha^2}}\exp\left(-\frac{\nu^2}{2\alpha^2}\right) \, .
\label{pg}
\end{equation}
The dependence on $\alpha$ is inspired by the fact that the range of interaction in the model under consideration  also affects the distribution of the fluctuations. In particular, in \eqref{pg} we place the standard deviation proportional to the range of interaction, that is $\alpha$.
The probability density \eqref{epdf} is thus reduced to 
\begin{equation}
\rho^*(\vec{z},\nu)=\sqrt{\frac{1}{2\pi\alpha^2}} \exp\left\{-\frac{\left[\nu+\alpha^2(\lambda_\Gamma-\lambda_E\bar{\psi^*})\right]^2}{2\alpha^2}\right\} \, .
\label{rhog}
\end{equation}
and the mean field equation to
\begin{equation}
-(-\Delta)^{\alpha/2}\bar{\psi^*}(\vec{z}) =\alpha^2(\lambda_E \bar{\psi^*}(\vec{z})- \lambda_\Gamma) \, .
\label{eq:fracmf}
\end{equation}

By means of \eqref{pg} and \eqref{rhog}, and neglecting the circulation constraint, the relative entropy functional can be written as
\begin{equation}
\mathcal{S}(\rho^*,\Pi_0)=-\frac{1}{\alpha^2}\mathcal{E}(\bar{q^*}) \, ,
\label{eq:SE}
\end{equation}
where $\mathcal{E}(\bar{q^*})$ represents the generalized potential enstrophy. Note that, beside the explicit dependence on $\alpha$ on the r.h.s., both $\rho^*$ and $\bar{q^*}$ have an implicit dependence on $\alpha$. 

From \eqref{eq:SE}, we can thus deduce that, for the  $\alpha$-models, there also  exists a \emph{selective decay principle} which can be stated as follow:
\emph{After a long time, the solutions of the $\alpha$-models of turbulence approach those states which minimize the generalized potential enstrophy for a given generalized energy.}

Note that, in a turbulent field, $\mathcal{E}$ shows a scale dependence which depends on the wavenumber of the instabilities and on $\alpha$ (e.g. \cite{pierrehumbertetal94,smith2002turbulent}), suggesting that the selective decay principle also follows a distinctive scale dependence which depends on the $\alpha$ model under consideration.

Setting  $\lambda_\Gamma$ to zero, the Lagrangian multiplier $\lambda_E$ in \eqref{eq:fracmf} is determined by the eigenvalues of the fractional Laplacian. Alternatively, since the stream function is determined up to a constant, $\lambda_{\Gamma}$ can be reabsorbed into $\bar{\psi^*}$ without loss of generality. For a double periodic domain of length $L=2\pi$,  the smallest eigenvalue is $\mu=\alpha^2\lambda_E=1$ and the stream function reduces to 
\begin{equation}
\psi(x,y)=a\sin(x)+b\sin(y)+c\cos(x)+b\cos(y),
\end{equation}
as for the $\alpha=2$ case \cite{majda_wang_2006}. So, when $a=A$, $b=c=d=0$ the solution corresponds to a simple unidirectional  shear flow, while when $a=b=A$, $c=d=0$ the solution corresponds to an array of $2$D swirling vortices.


\remark{Different prior distributions lead to different mean field equations, which can also be nonlinear \cite{Pasmanter1994,Gurarie2004,Bouchette2012}. Although the Gaussian prior distribution generates a quadratic entropy functional, suggesting thus the extension of selective decay principle also for all the $\alpha$-models, nonlinear mean field equations can be particularly important for geophysical flow \cite{bouchet_sommeria_2002,Dibattista}.  Moreover, a nonlinear prior distribution can give rise to a higher degree polynomial entropy functional, that can thus be used to explain flow transitions \cite{Bouchet2009,Bouchette2012}. }

\section{\label{sec:level4}Numerical Investigations}
In the previous section, the existence of a selective decay principle was suggested by a probability density obtained with a particular
prior distribution. It is however not proved that this principle holds in general. In the following section, we will explore numerically the evolution of a turbulent flow for different values of $\alpha$ in order to investigate the existence of the selective decay principle for our i.c.s. 

We perform numerical simulations using a semi-spectral scheme in a double periodic domain of length $L=2\pi$. The Jacobian is computed by means of the Arakawa \cite{ARAKAWA1966119} discretization. Time integration is performed with a fourth order explicit Runge-Kutta scheme. To avoid the accumulation of enstrophy at small scales, we  multiply each Fourier mode by an exponential function of the form $f(w)=\exp(-a w^m)$ where $w$ is a variable proportional to the meridional or zonal wavenumbers \cite{HOU2007379,constantin2012new,ragone2016study}. 

All the simulations are carried out in a fixed geometry with an aspect ratio of  one. As explained in \cite{Bouchette2012} the geometry of the domain plays a crucial role for the formation of structures  at the equilibrium, with elongating domain favourizing bar-states, thus unidirectional flows with a functional relation between $q$ and $\psi$ that is tanh-like. Another important key parameter that guides the bifurcation diagram for the equilibrium structures is the initial distribution for the different fields. In fact global distributions, such as bimodal distributions, that lead to tanh-like $\psi-q$ function relation, will favor unidirectional flow states, while initial distributions  leading to sinh-like relations, such as dilute vorticity leaks, will favor the formation of dipolar states \cite{majda_wang_2006,Pasmanter1994}.

Within a periodic square, the extrema are degenerate and both unidirectional or dipolar structures of the flow can emerge.
This property of the domain is interesting because, although the initial global distribution here considered is bimodal, the different modes of the flow can compete with each other during the relaxation phase. This competition can be explored with the natural order parameter
\begin{equation}
\gamma=\dashint_{\Omega} q e^{iy}\,{\rm{d}}\vec{z}.
\label{gamma}
\end{equation}
For an unidirectional flow in a squared domain,  $|\gamma|$ will tend to zero since the active scalar at the equilibrium will be proportional to one of the eigenstates of the fractional Laplacian in the square domain, while it will be different from zero if the solution is a superposition of two eigenstates. This is an empirical way to investigate the equilibrium phase transitions, taking into consideration the out of equilibrium topological transitions of the flow. The latter were observed for the Navier-Stokes equations subjected to stochastic forcing \cite{Bouchet2009,Bouchette2012}. In our case no forcing is used, but the turbulence generated by the filament breaking  starting from  hyperbolic saddle i.c.s,  and the associated numerical dissipation, are sufficient to induce this type of transitions.

In order to determine the extrema suggested by the statistical theory and in order to compare the theory with the simulations, we adopt an iterative algorithm proposed by Turkington and Whitaker \cite{Turkington1996}.  The main idea behind the algorithm is to consider simpler problems that arise from the linearization of the energy constraint and that, in the limit, converge to the solution sought \cite{majda_wang_2006}. To simplify we will consider only the energy and the normalization constraints. 

\subsection{\label{sec:level42} Hyperbolic Saddle}

We explore the selective decay and the equilibrium extrema associated to a hyperbolic saddle geometry. These i.c.s have aroused a certain interest since early studies have suggested that it could be able to develop singularities for the inviscid $\alpha=1$ (SQG) case \cite{constantinetal94,constantin1994singular,constantin1998nonsingular,constantin2012new}. 
The possible existence of singularities in SQG is suggested by considering the horizontal gradient of \eqref{eq1}, which yields 
\begin{equation}
\frac{D \nabla^\perp q}{D t}  = \nabla \vec{u} \cdot \nabla^\perp q  \, ,
\label{eq:grad sqg}
\end{equation}
where $D / Dt$ represents the material derivative. The resulting equation is a $2$D counterpart of the $3$D Euler equation, with $\nabla q \leftrightarrow \vec{\omega}_{3D}$ and with an emerging $2$D stretching term on the r.h.s which might be responsible for the blow-up. Note  that, for $\alpha=2$, the r.h.s. of \eqref{eq:grad sqg} is zero and no blow-up occurs. The finite time blow-up criteria of the resulting equation is the correspondent of a Beale-Kato-Majda criteria \cite{beale1984remarks}, which in the case of SQG reads as \cite{constantinetal94} 
\begin{equation}
\int_{0}^{T} || \nabla q ||_{L^{\infty}} (s) ds \rightarrow \infty \, \text{as} \, T \rightarrow T^* \, . 
\label{eq:bkm}
\end{equation}

Successive studies have ruled out the formation of singularities in the specific case of the hyperbolic saddle geometry \cite{cordoba1998nonexistence}, with $\nabla q \propto \exp \exp t$. Other studies \cite{pierrehumbertetal94,scott2011scenario,scott2014numerical,scott2019scale} have however suggested that during its time evolution, the hyperbolic saddle forms a filament which undergoes secondary instabilities and  these instabilities can lead to a finite time singularity via a self-similar cascade of $\nabla q$ toward smaller and smaller scales.
 
In the following we will compare the evolution of the hyperbolic saddle geometry for the cases $\alpha=1$ and $\alpha=2$.  
The initial condition is given by
\begin{equation}
q(\vec{z})=\sin(x)\sin(y)+\cos(y),
\label{sic}
\end{equation}
for both $\alpha=1$ and $\alpha=2$. 
Note that since we prescribe the same initial vorticity field \eqref{sic} for both the simulations  but $\psi$ changes according to \eqref{invp}. Changing the interaction's range has the same effect of changing the initial conditions.
Simulations are performed with increasing grid resolutions $512 \times 512$,  $1024 \times 1024$, $2048 \times 2048$ and $4096 \times 4096$. 

We simulate the flow until $T=16$. For the low resolution case of $512 \times 512$, we simulate the flow until $T=5 \times 10^{3}$. All simulations are performed with time step $\Delta t=10^{-3}$.

The  dissipation  is given only by the exponential filter. 
The parameters for the filter in the Fourier space are set as in \cite{constantin2012new}, that is $a=36$, $m=19$. 

\begin{figure}[hbt!]
\begin{center}
\subfloat[$\alpha=1$]{
{\includegraphics[width=.5\textwidth]{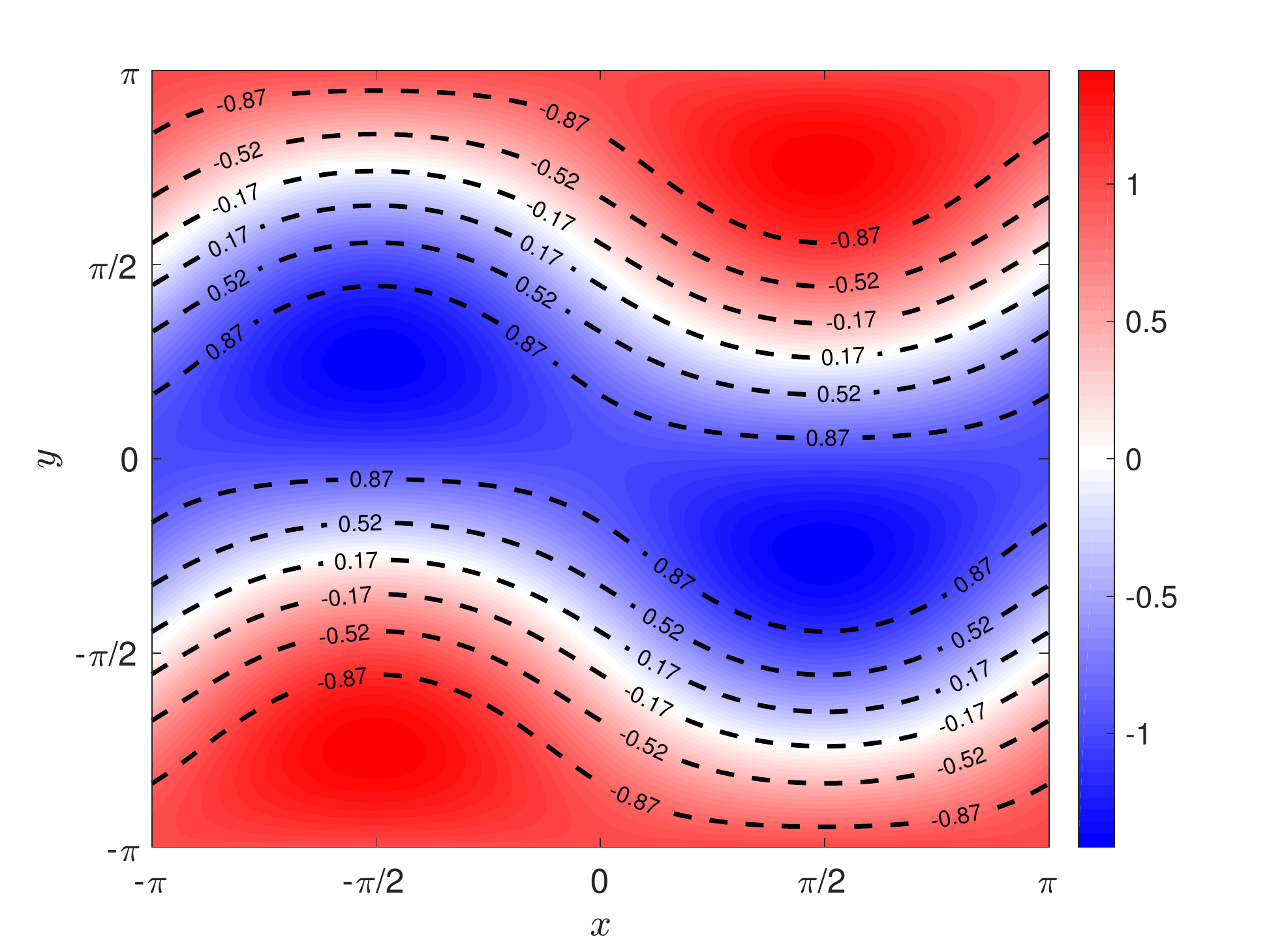}}}%
\subfloat[]{
{\includegraphics[width=.5\textwidth]{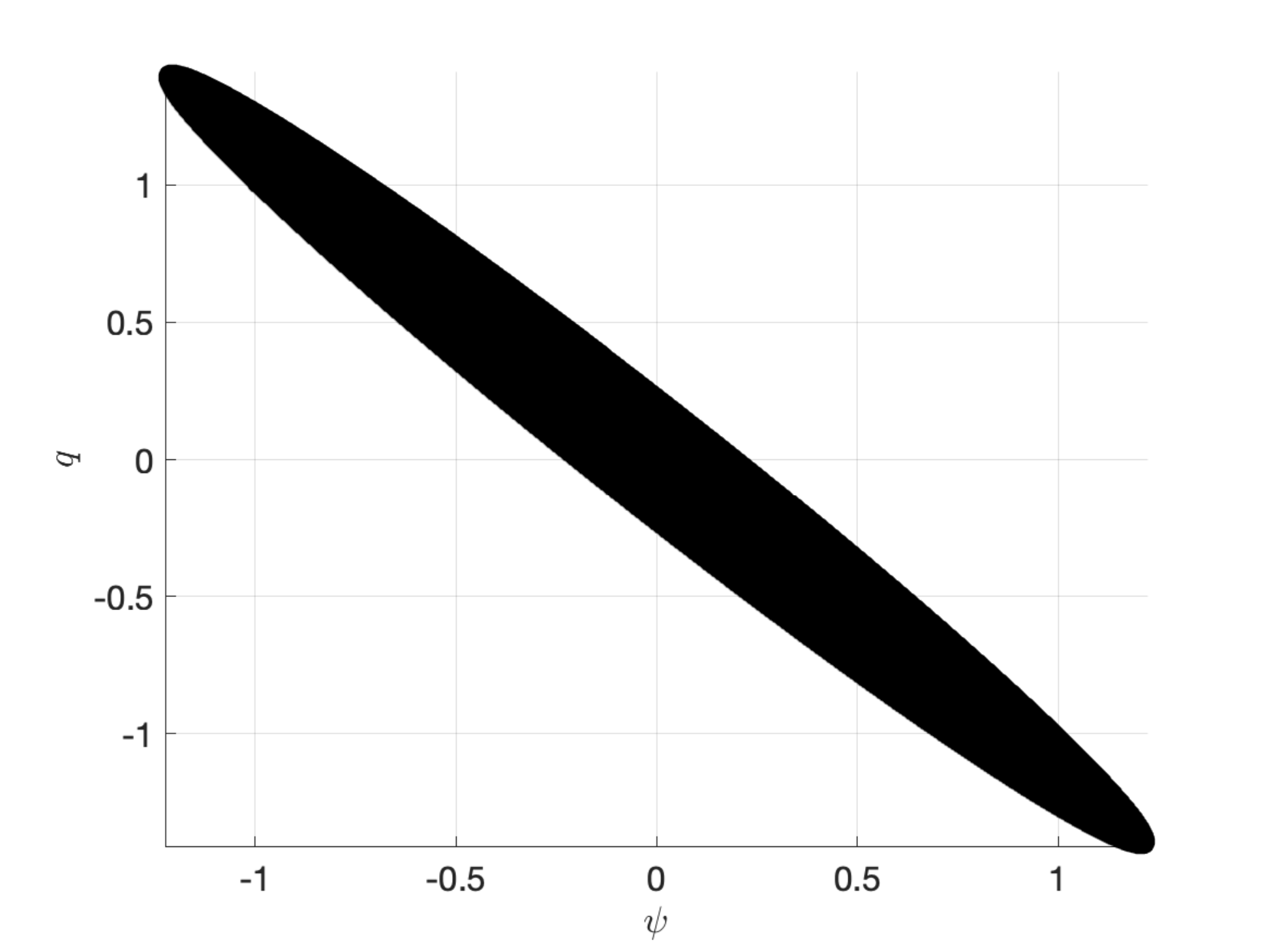}}}\\%
\caption{Initial conditions \eqref{sic}. The grid resolution here is  $4096\times4096$ grid points. The left panel shows the $q$ field with the superimposed stream function (dashed line). The right panel shows the corresponding $\psi-q$ relation for the case $\alpha=1$.}%
\label{fig:n4}
\end{center}
\end{figure}

Figure \ref{fig:n4} shows the initial active scalar field with the correspondent stream function (dashed line) in the left panel for $\alpha=1$ and, the  functional relation $\psi - q$ on the right.

\begin{figure}[hbt!]
\begin{center}
\subfloat[$T=8$]{
{\includegraphics[width=.5\textwidth]{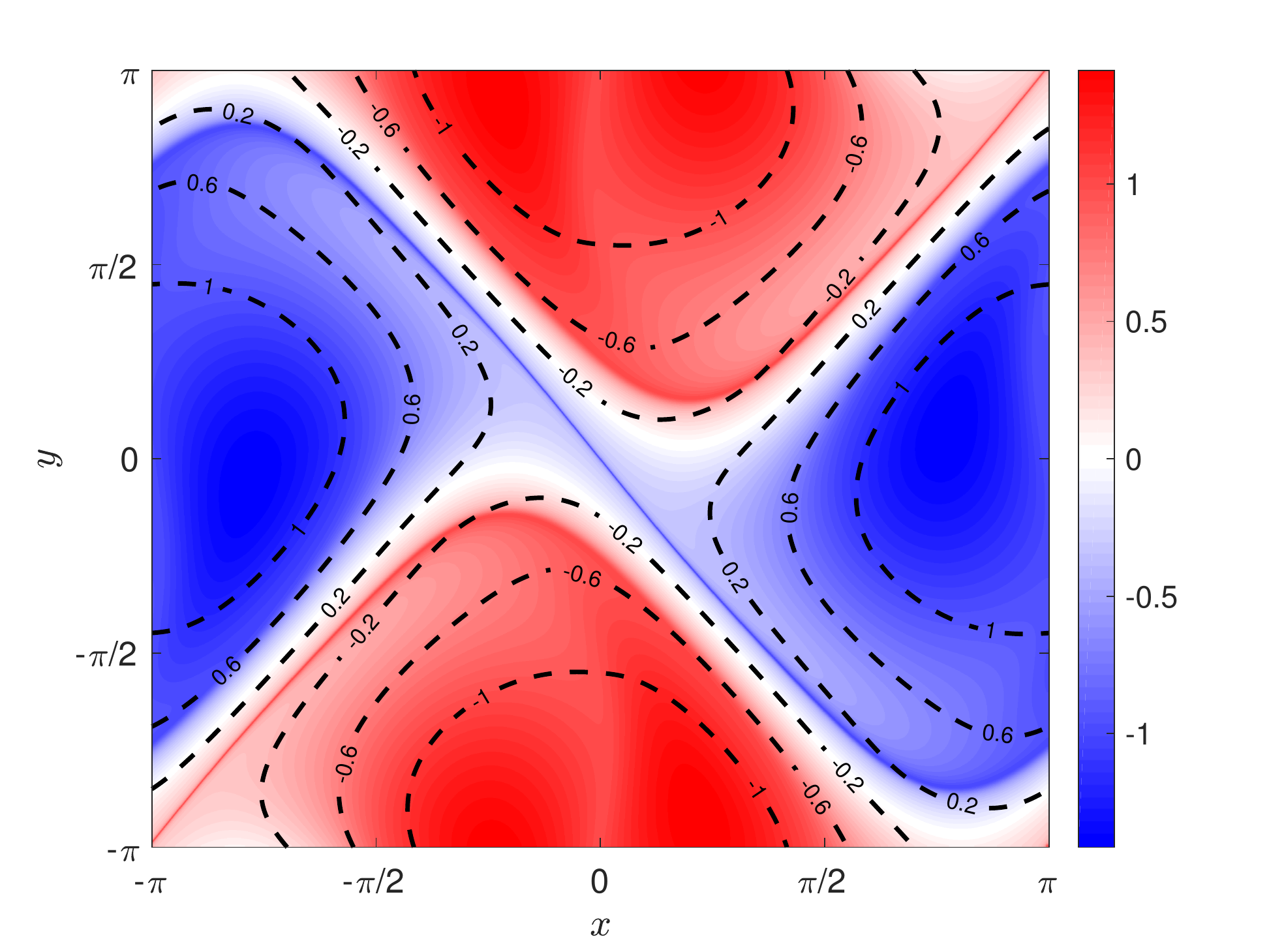}}}%
\subfloat[]{
{\includegraphics[width=.5\textwidth]{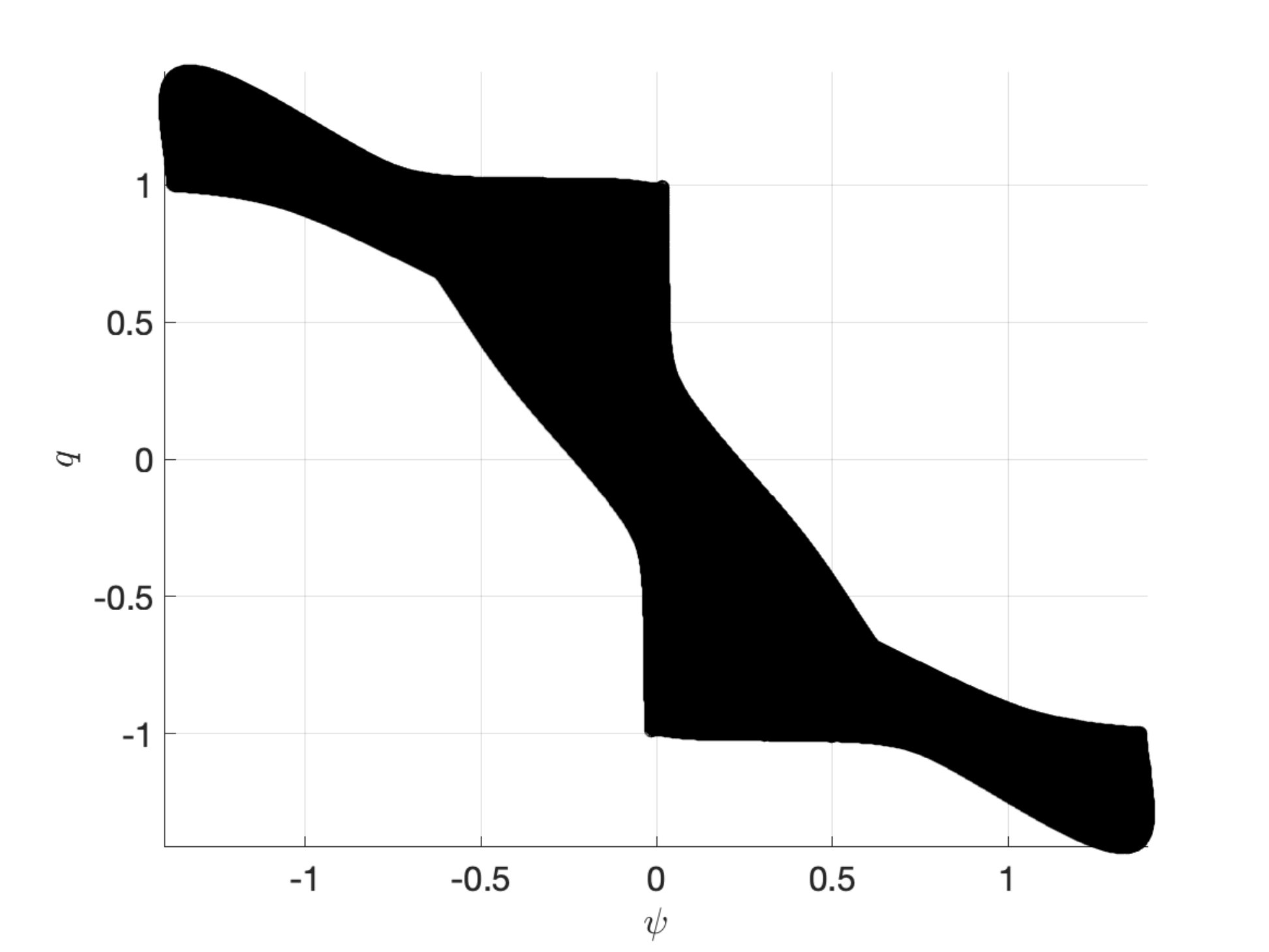}}}\\%
\subfloat[$T=11.5$]{
{\includegraphics[width=.5\textwidth]{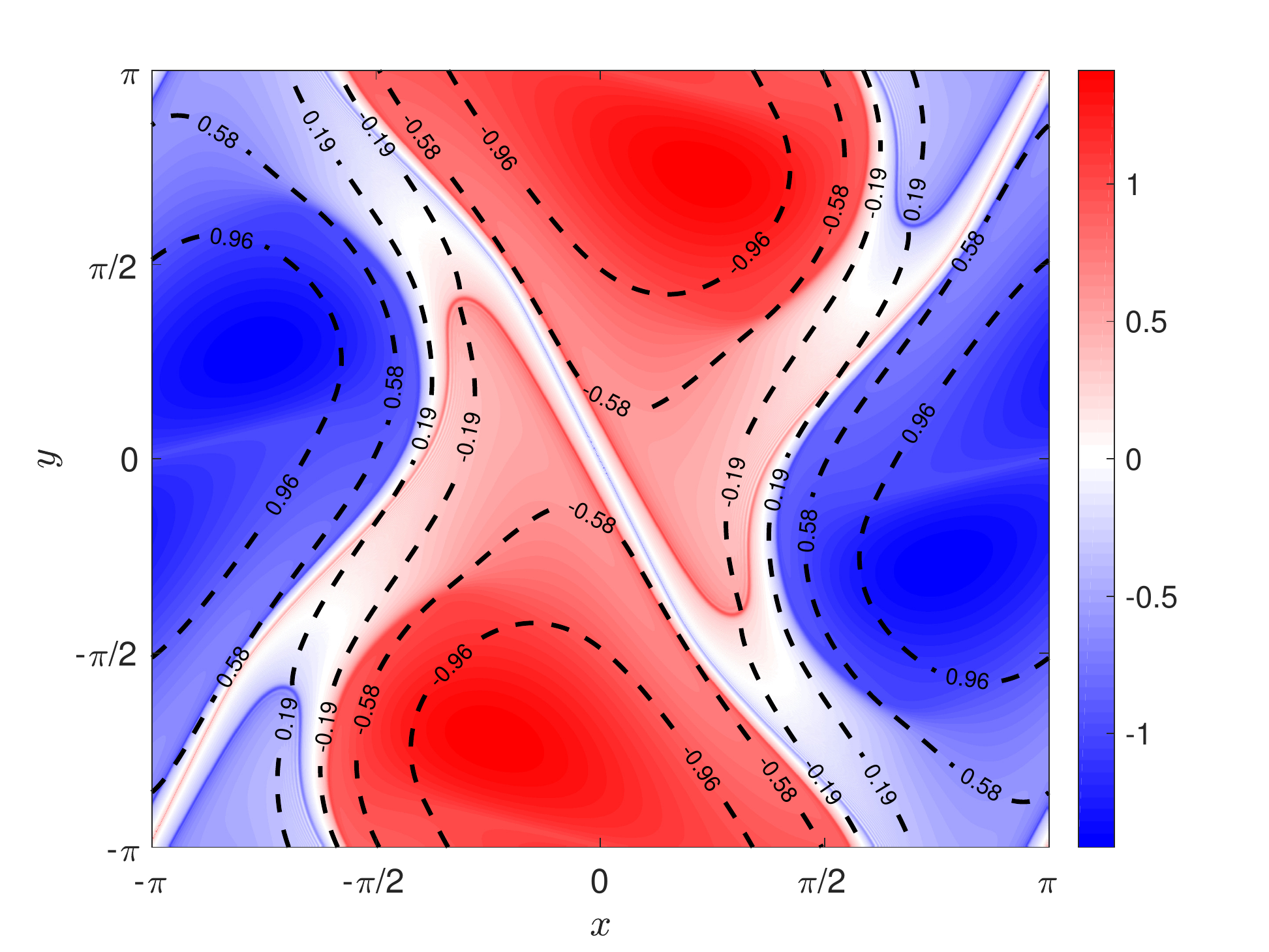}}}%
\subfloat[]{
{\includegraphics[width=.5\textwidth]{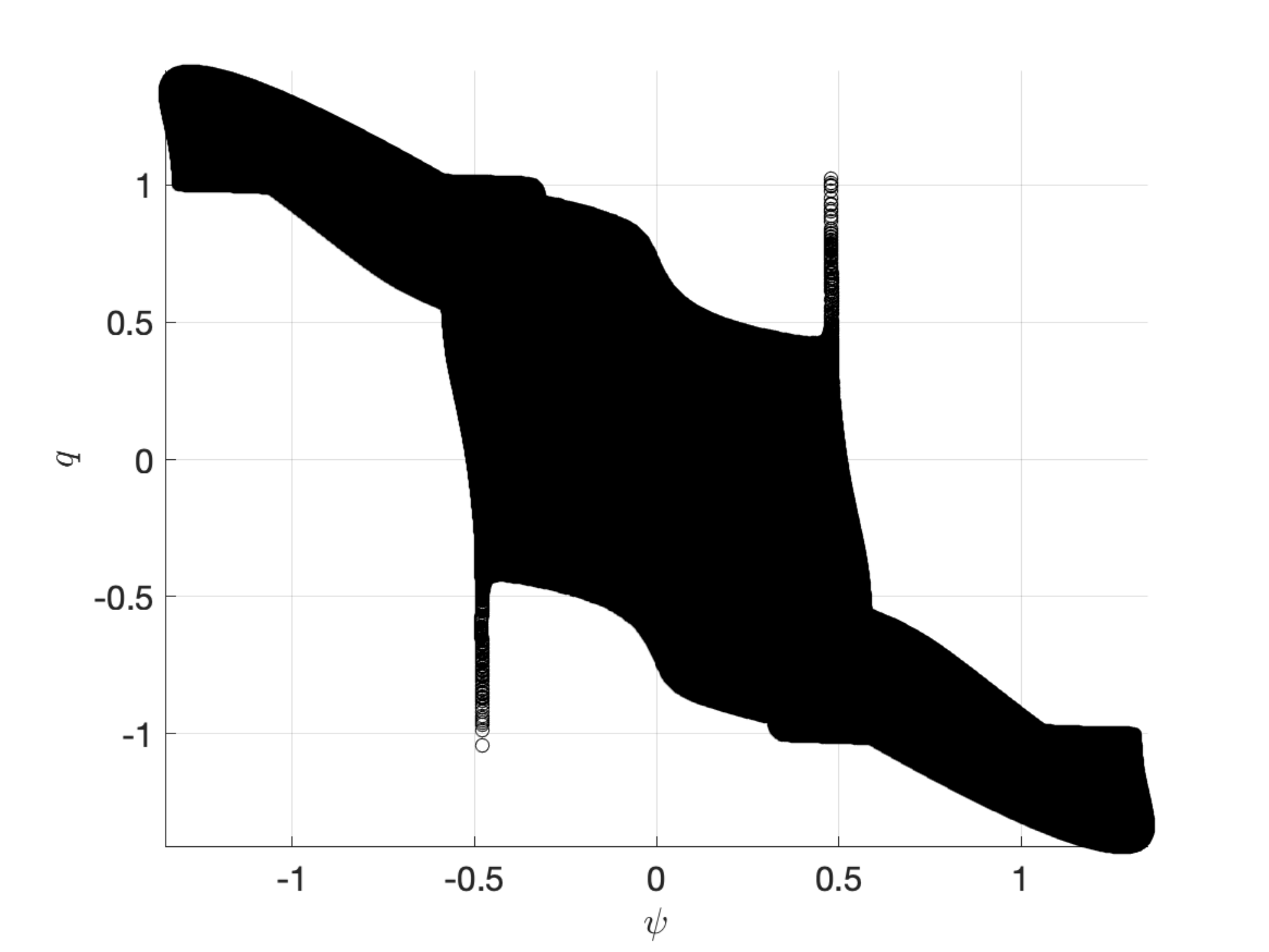}}}\\%
\subfloat[$T=15.9$]{
{\includegraphics[width=.5\textwidth]{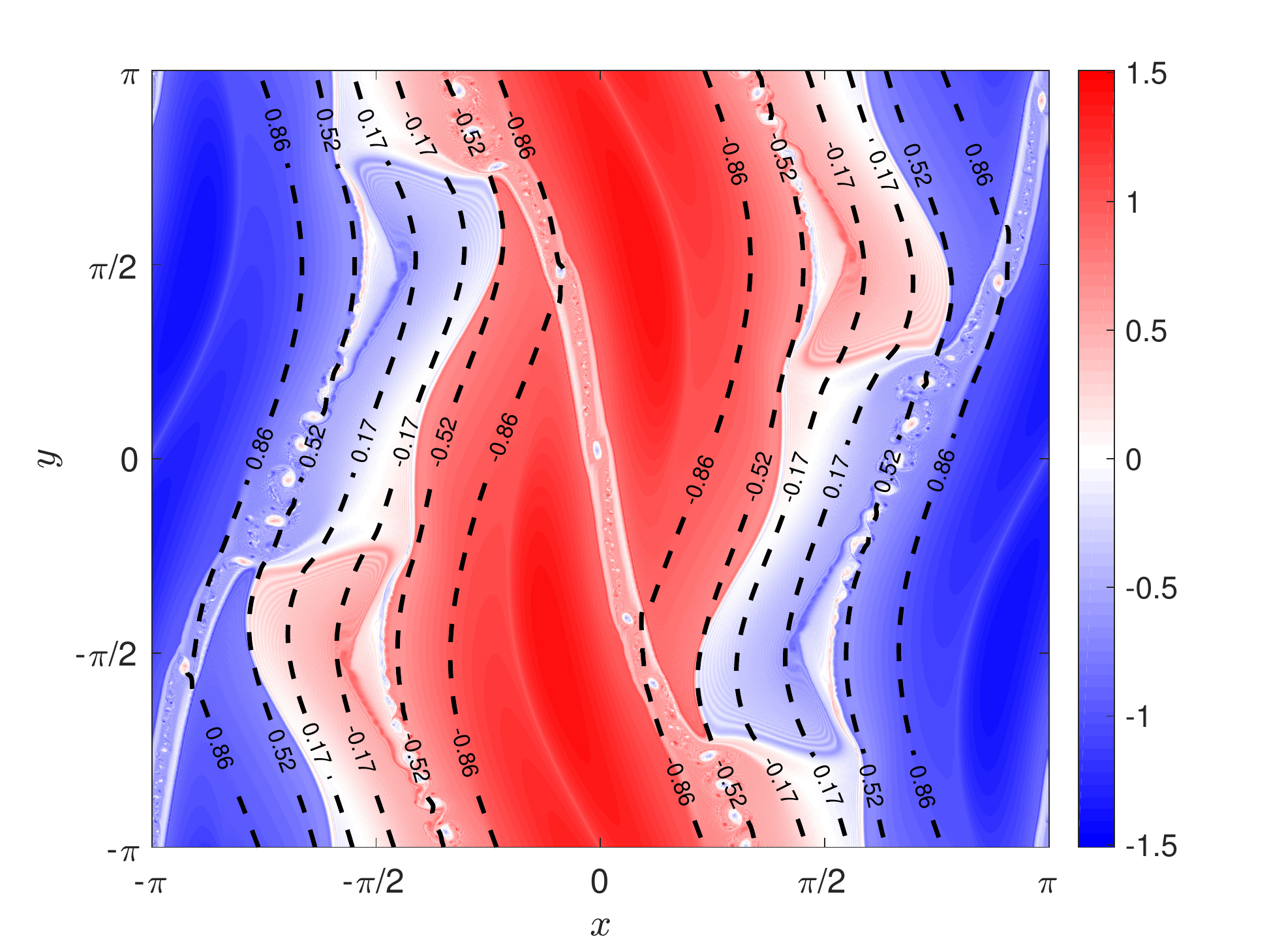}}}%
\subfloat[]{
{\includegraphics[width=.5\textwidth]{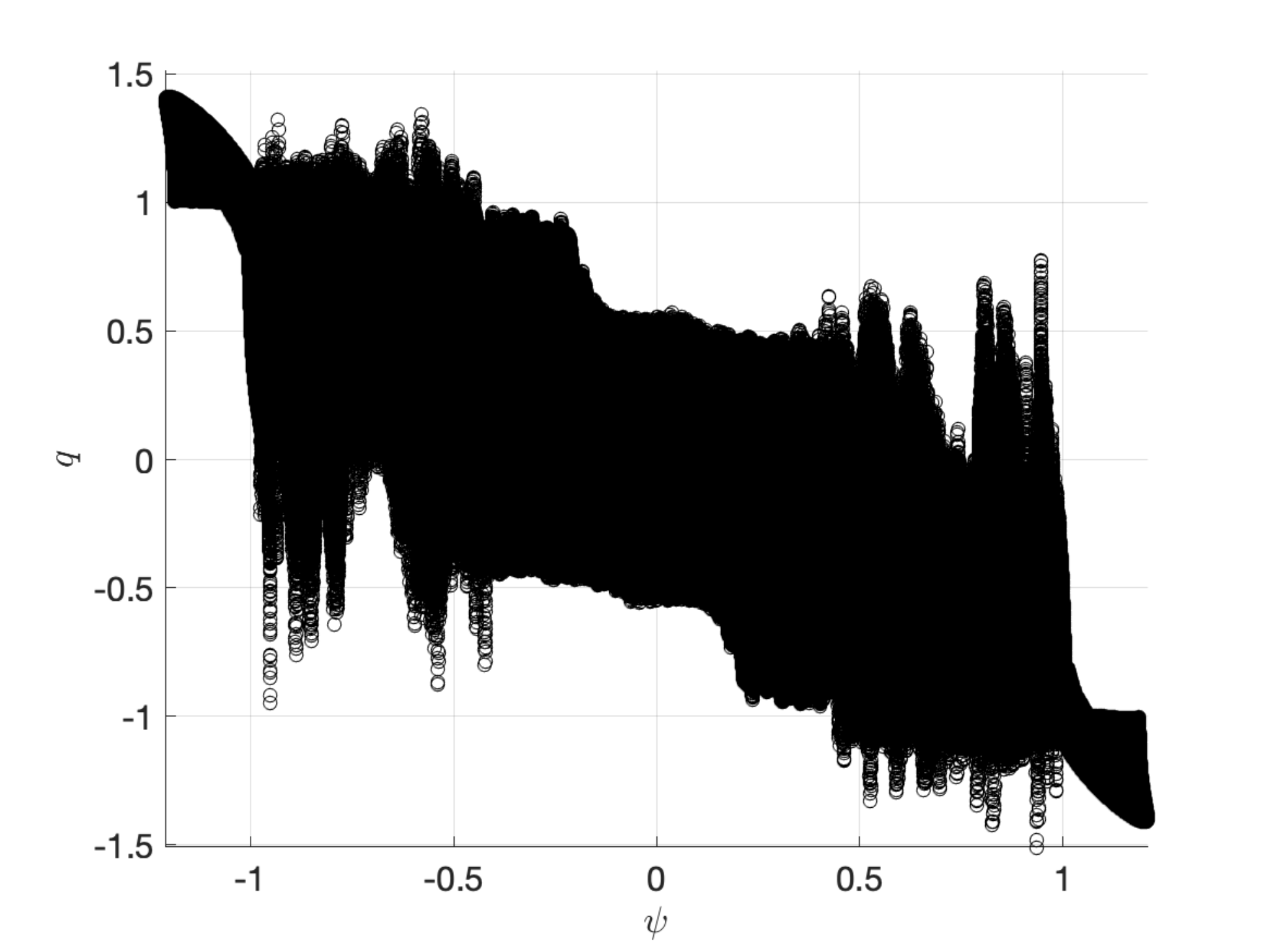}}}\\%
\caption{Left panels: $q$ (coloured) and $\psi$ (dashed lines) fields for the case $\alpha=1$ (a) during the front intensification at $T=8$, (c) during the front deformation at $T=11.5$  and (e) after the disruption of the front at $T=15.9$. Right panels: scatter plots of $q$  versus $\psi$ for the corresponding three times. The grid resolution is of $4096\times4096$. }%
\label{fig:n5}
\end{center}
\end{figure}

\begin{figure}[hbt!]
\begin{center}
{\includegraphics[width=.8\textwidth]{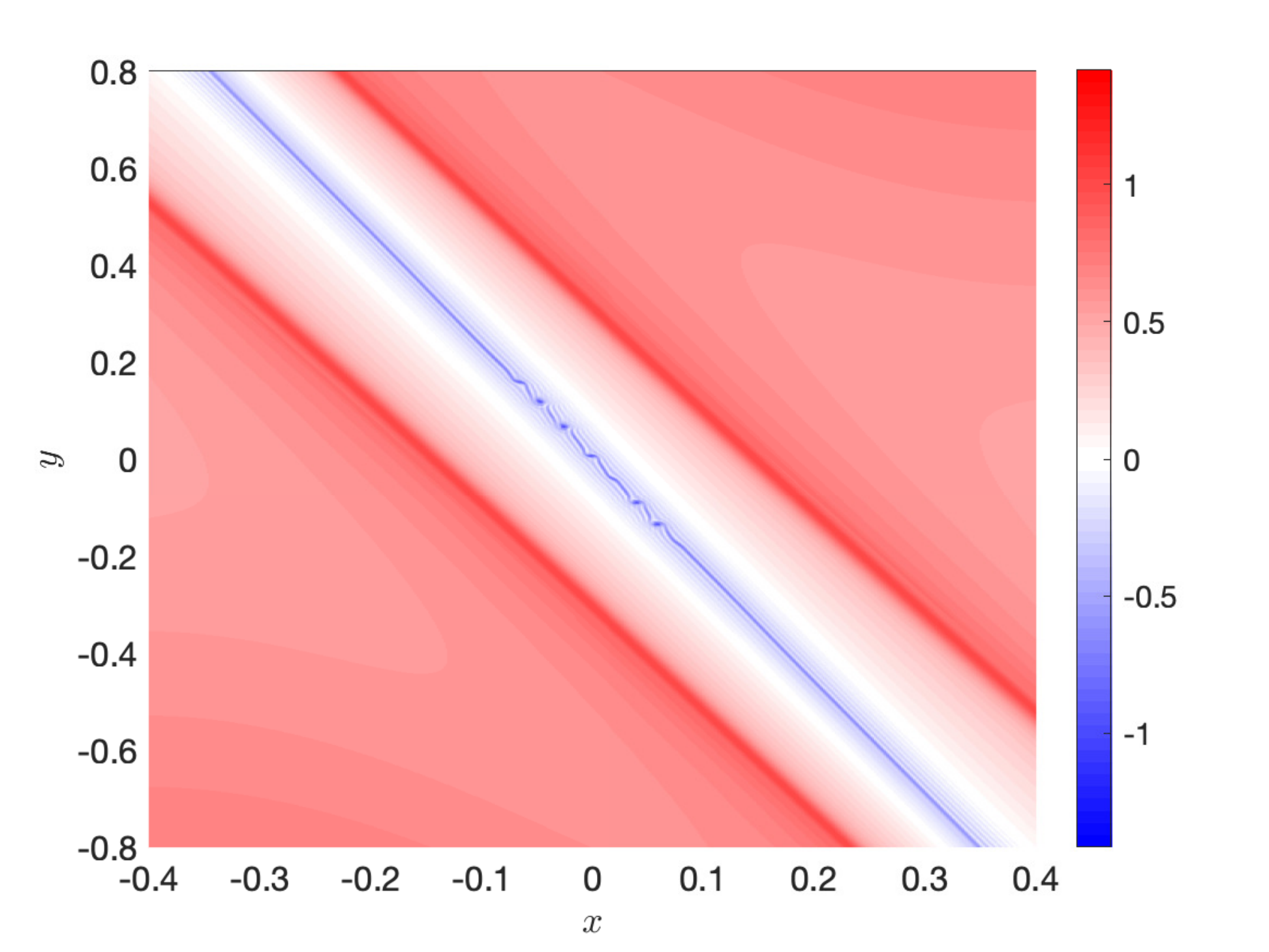}}%
\caption{ Central region of Fig. \ref{fig:n5}c at $T=11.5$, showing the deformation of the front by means of secondary instabilities .
}%
\label{fig:n6}
\end{center}
\end{figure}

Figure \ref{fig:n5} shows the evolution of the flow, which consists in an intensification of the front at $T=8$, a breaking down of the filament by means of secondary instabilities at $T=11.5$ and a disruption of the central filament at $T=15.9$.   

The scatter plots of $q$  versus $\psi$ on the right panels of figure \ref{fig:n5} highlight the nonlinear relations out of the equilibrium between the active scalar and the stream function during the three stages of the development of the flow.  

The form of the secondary instabilities, which are responsible for the breaking of the central filament, is shown in Figure \ref{fig:n6}. As noted by \cite{scott2011scenario}, even if the simulations are able to resolve the secondary filament connecting these secondary instabilities, even for the highest resolution simulation the formation of these happens when the evolution of the flow is past its inviscid stage. The study of the breaking of the central filament at different resolutions will, however, give us hints about the possibility to study this phenomena making use of the selective dissipation.

\begin{figure}[hbt!]
\begin{center}
{\includegraphics[width=.8\textwidth]{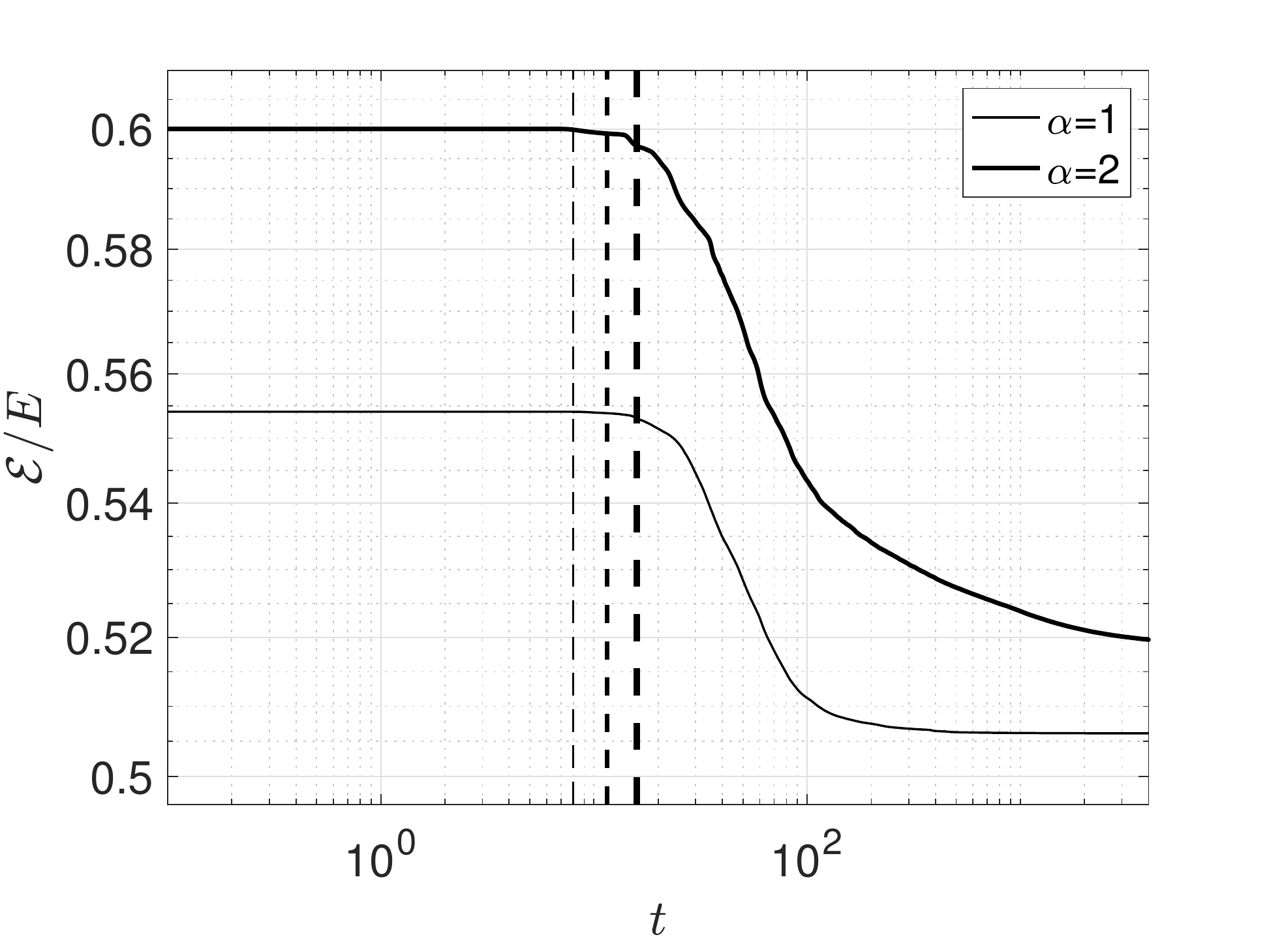}}%
\caption{The $R$ ratio for the simulations with initial condition \eqref{sic}. The vertical dashed lines represent the times $T=8$, $T=11.5$ and $T=15.9$ respectively.}%
\label{fig:n7}
\end{center}
\end{figure}

Figure  \ref{fig:n7} shows the behaviour in time of the ratio between  the generalized enstrophy and energy, $R=\mathcal{E}/E$, which has been used to express in a mathematical rigorous way the selective decay principle for $\alpha=2$ \cite{majda_wang_2006,Venaille2015}. It should be noted that for $\alpha=2$, the quantity $P=(2\pi/L)^\alpha E/\mathcal{E}$, that for our simulations reduce to $P=R^{-1}$,  has been observed to be $P\approx1$ when a general  unstable conditions relaxes toward an equilibrium state \cite{TABELING20021}. 
The trend of the ratio $R$ shows that, even at the end of the integrations, the simulations do not reach a true steady state. At $T=5 \times 10^3$ the solutions still exhibit oscillations and the ratio $R$ is not completely equilibrated \cite{Ohkitani_2010,Ohkitani_2012}. This quantity also says that the generalized enstrophy, according to the selective decay, decreases in time also for the case $\alpha=1$ and that the depletion is faster than the one of the generalized energy.

\begin{figure}[hbt!]
\begin{center}
\subfloat[$\alpha=1$]{
{\includegraphics[width=.5\textwidth]{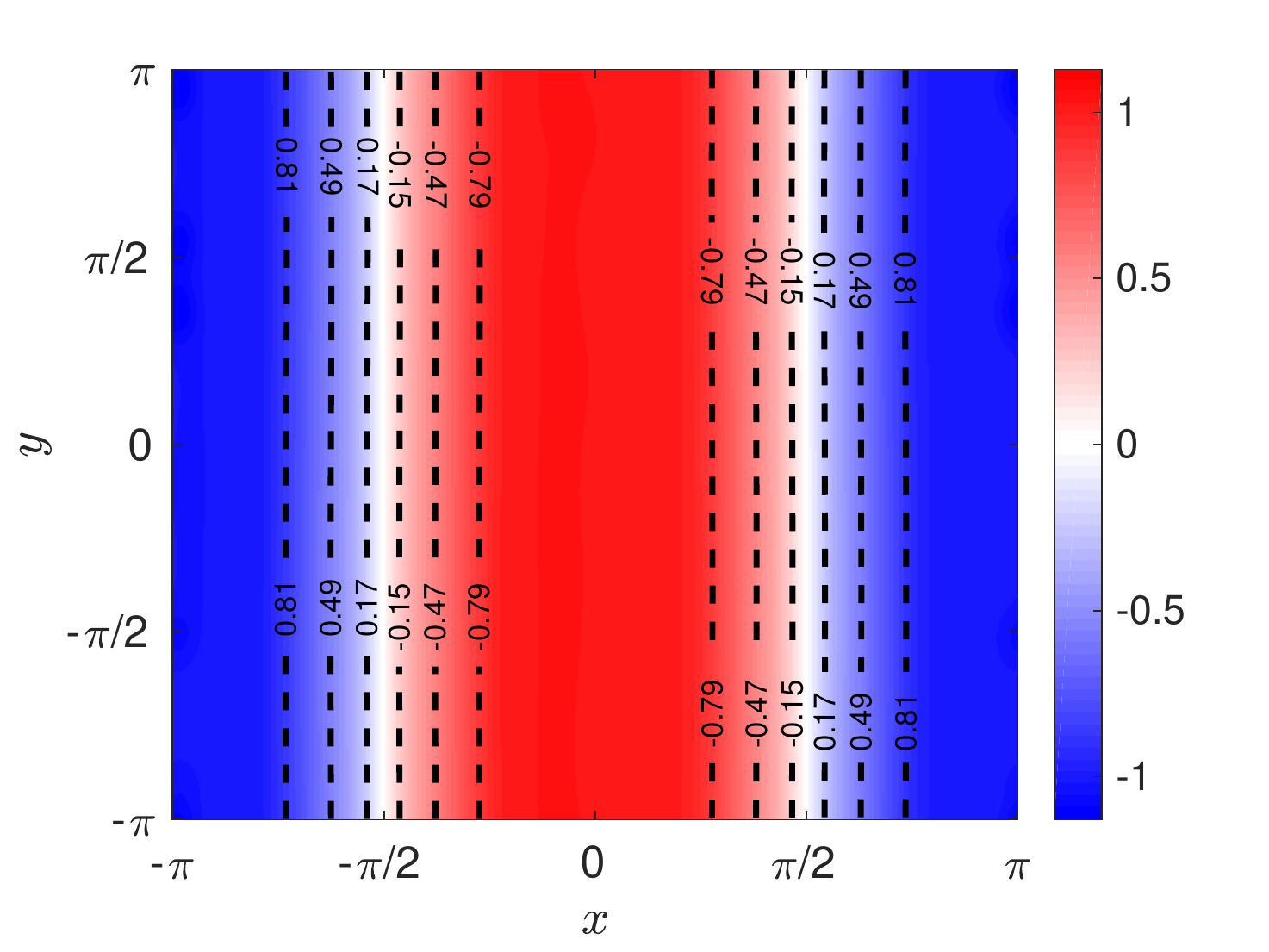}}}%
\subfloat[]{
{\includegraphics[width=.5\textwidth]{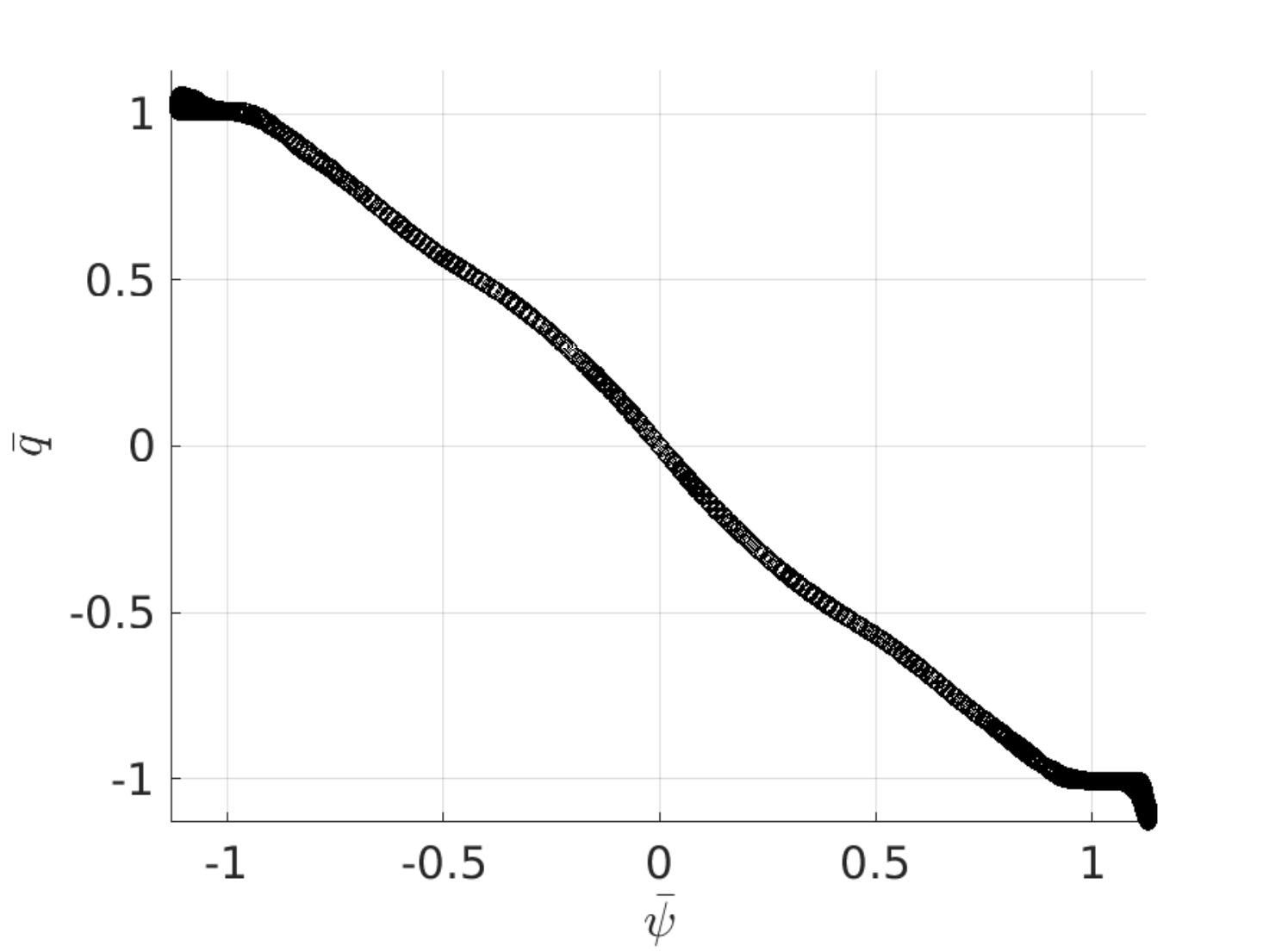}}}\\%
\subfloat[$\alpha=2$]{
{\includegraphics[width=.5\textwidth]{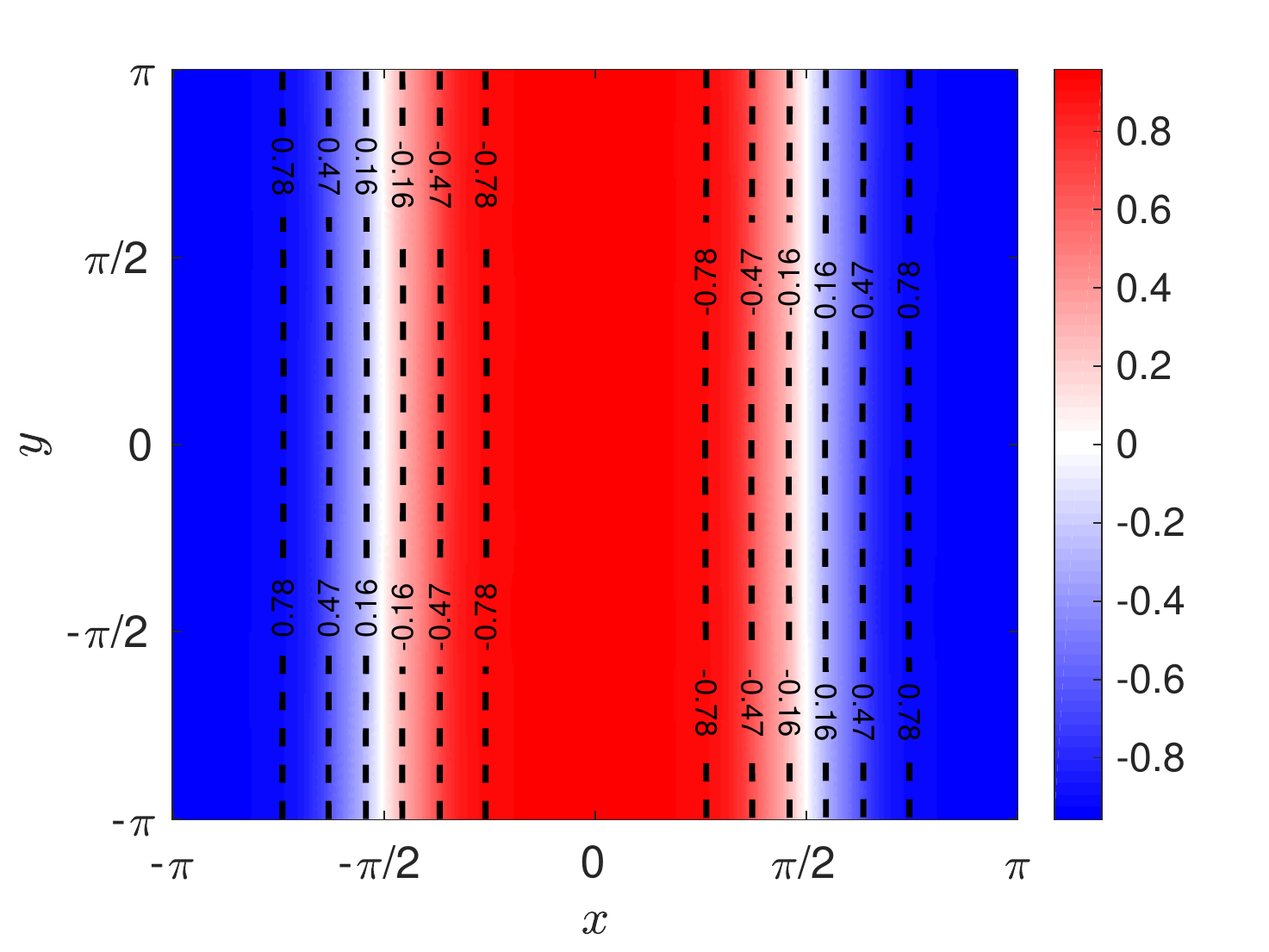}}}%
\subfloat[]{
{\includegraphics[width=.5\textwidth]{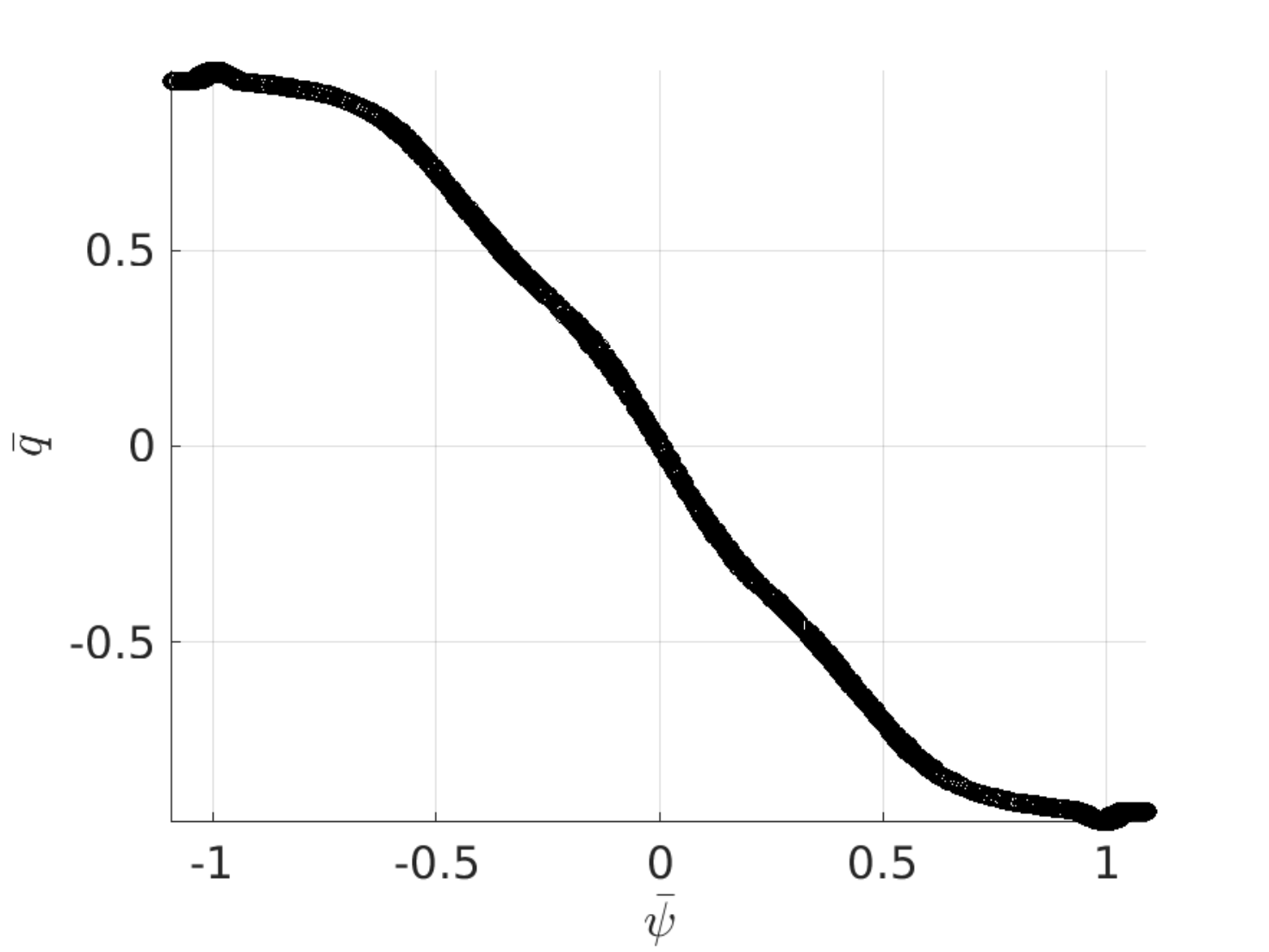}}}\\%
\caption{Left panels: time average of  $q$ and $\psi$ fields over the last $\sim80$ eddy turnover times. Right panels: functional relation between the same two fields for the case (a,b) $\alpha=1$, and (c,d) $\alpha=2$.  For this long run  the grid resolution used is set to  $512\times512$ grid points.}%
\label{fig:n8}
\end{center}
\end{figure}

To study final quasi steady state, we consider a time average of the fields on the last $\sim80$ eddy turnover times, here defined as $T_{eddy}=(2\pi)^{2-\alpha} /q_{rms}$ (Figure \ref{fig:n8}).  This allows us to remove the fluctuations about the quasi  steady state. For both $\alpha=1,2$, the flow evolves toward a meridionally oriented structure. The comparison between the two cases also shows  that for $\alpha=1$ the functional relation $\psi - q$ is approximately linear, while for $\alpha=2$ the nonlinearity is intensified.
\begin{figure}[hbt!]
\begin{center}
{\includegraphics[width=.6\textwidth]{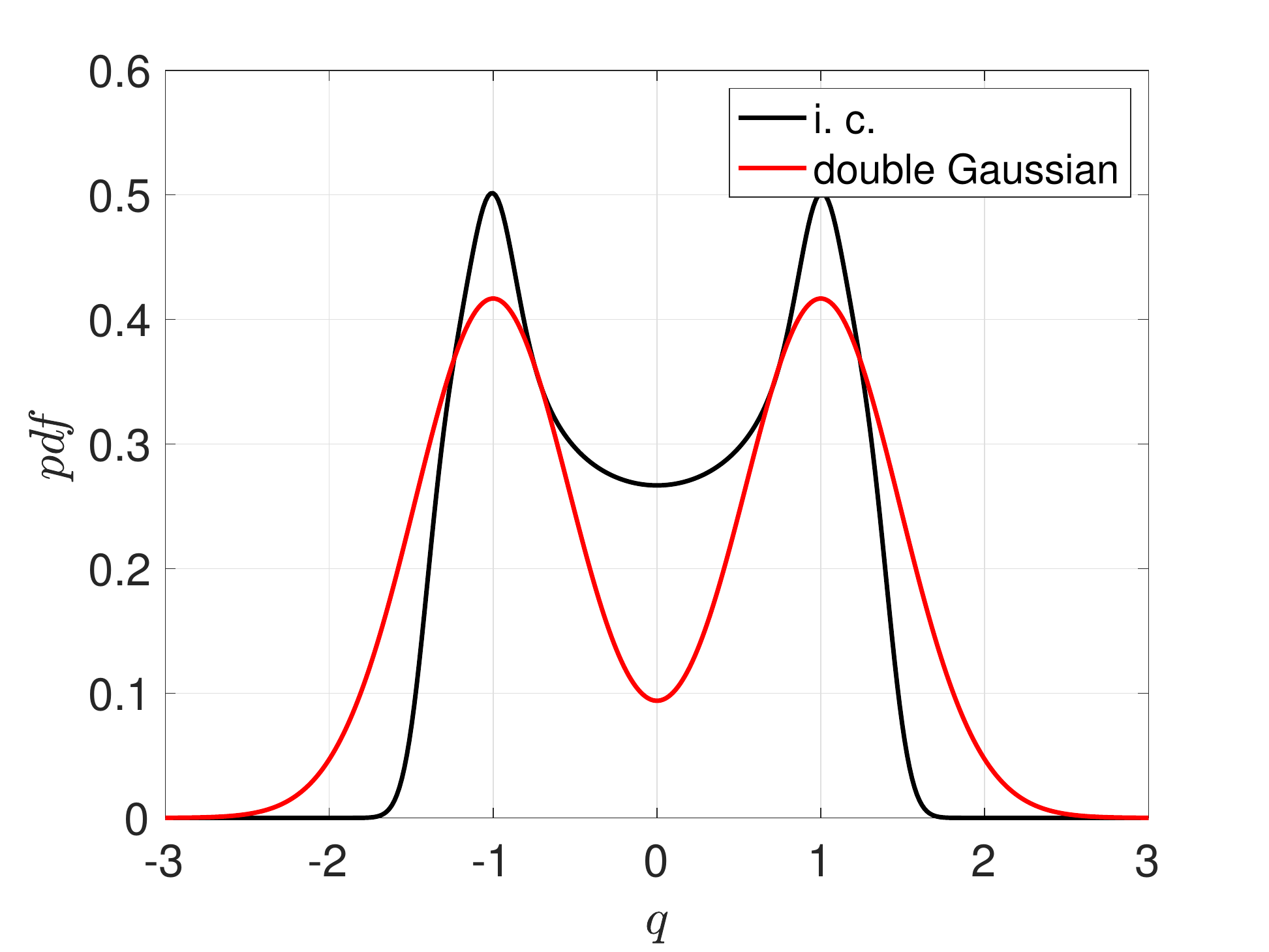}}
\caption{ Initial global probability density function when $q$ is given by \eqref{sic}, black line, and a bimodal global distribution obtained by the sum of two Gaussian distribution, red line. }%
\label{fig:singpdf}
\end{center}
\end{figure}
For the bimodal initial global distribution of vorticity generated by the i.c.s under consideration (Fig. \ref{fig:singpdf}, black line), we expect an equilibrium state that is unidirectional and characterized by a tanh-like functional relation between vorticity and stream function. This can be easily shown considering, for instance, a bimodal prior distribution approximated  by the sum of two Gaussian (Fig. \ref{fig:singpdf}, red line),
\begin{equation}
\Pi_0(\nu)=\frac{1}{2\sigma\sqrt{2\pi}}\left[ \exp \left(-\frac{(\nu+1)^2}{2\sigma^2}\right)+\exp \left(-\frac{(\nu-1)^2}{2\sigma^2}\right) \right].
\label{eq2gauss}
\end{equation}
The value of $\sigma$ is chosen using a best fit procedure between $\Pi_0$ and the distribution. 
  From \eqref{eq2gauss} and \eqref{mfe} we obtain
\begin{equation}
\bar{q^*}=\lambda_E \sigma^2 \bar{\psi^*} + \tanh(\lambda_E\bar{\psi^*}),
\label{tanhqp}
\end{equation}
as mean field equation for the equilibrium state. Although the initial global condition is not Gaussian, the principle of selective decay seems to hold. 
\begin{figure}[hbt!]
\begin{center}
\subfloat[$\alpha=1$]{
{\includegraphics[width=.5\textwidth]{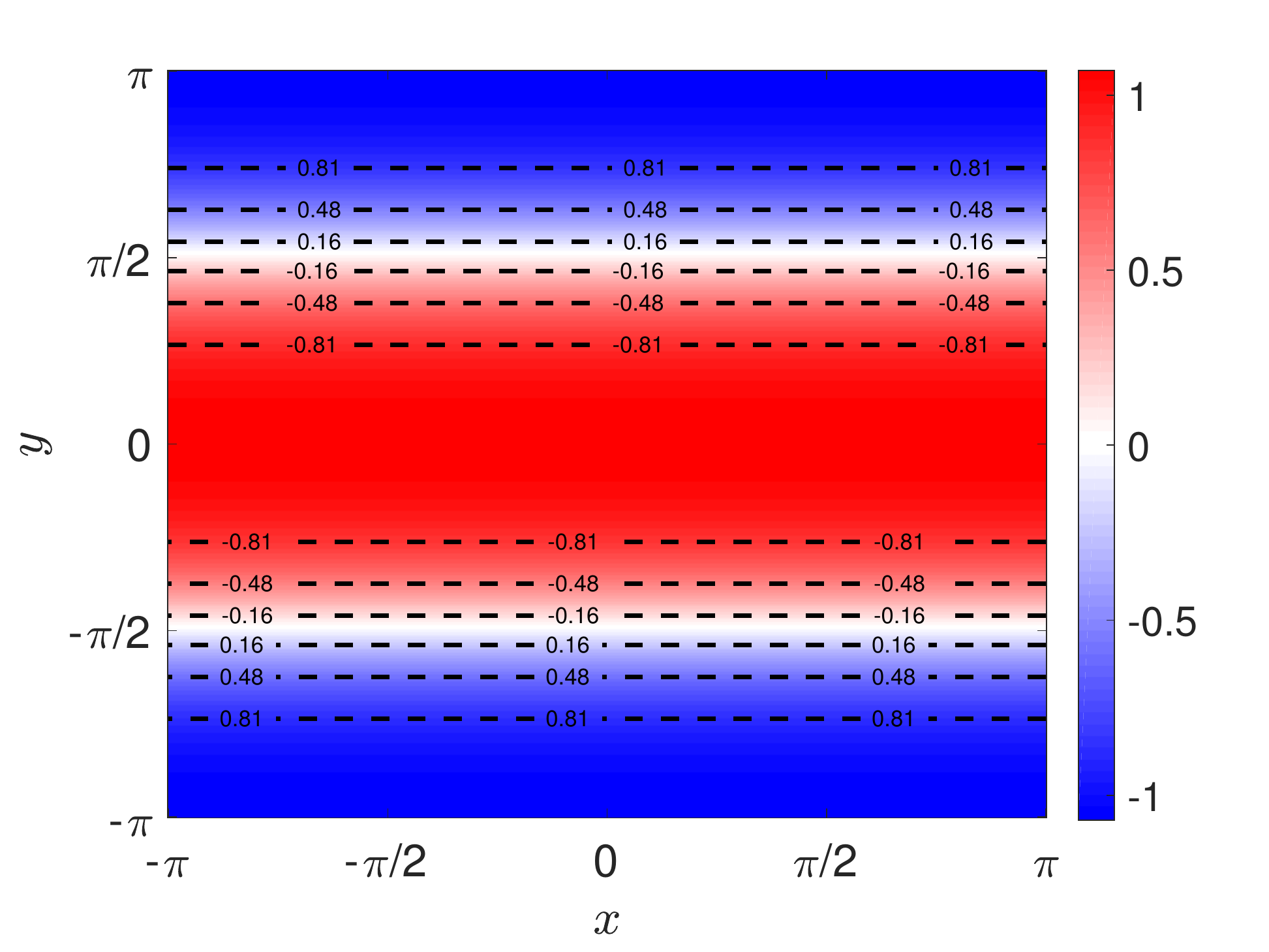}}}%
\subfloat[]{
{\includegraphics[width=.5\textwidth]{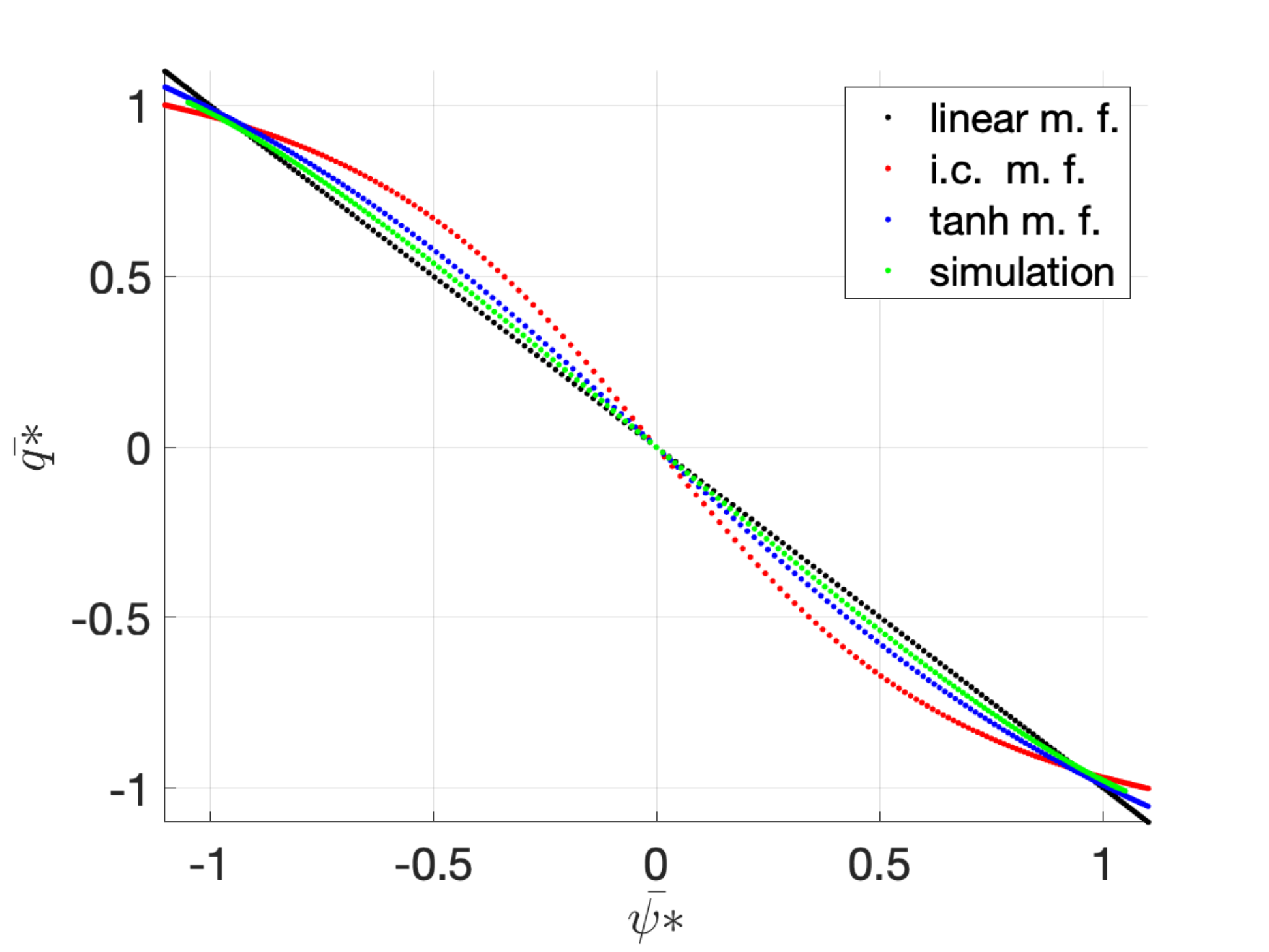}}}\\%
\subfloat[$\alpha=2$]{
{\includegraphics[width=.5\textwidth]{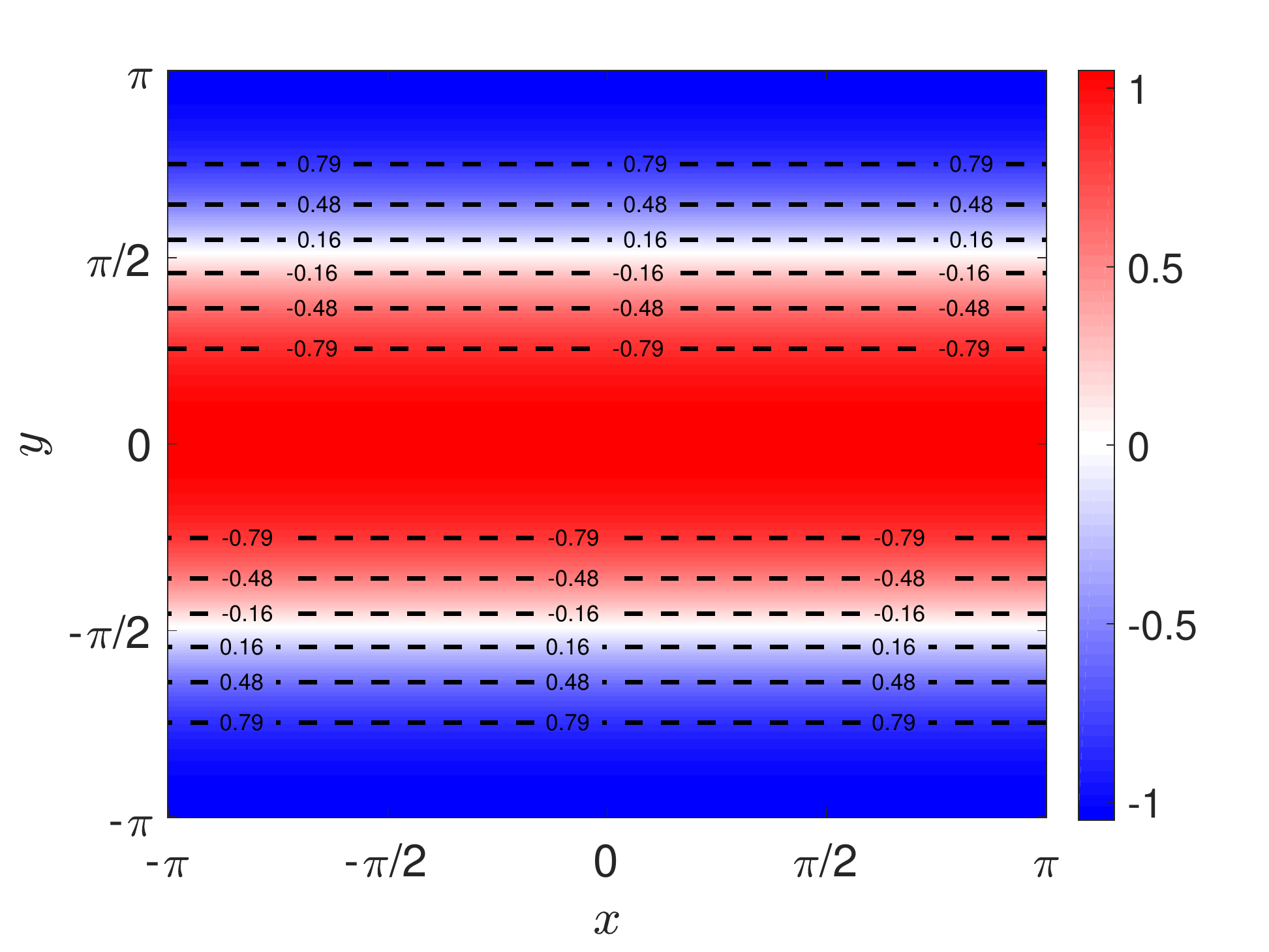}}}%
\subfloat[]{
{\includegraphics[width=.5\textwidth]{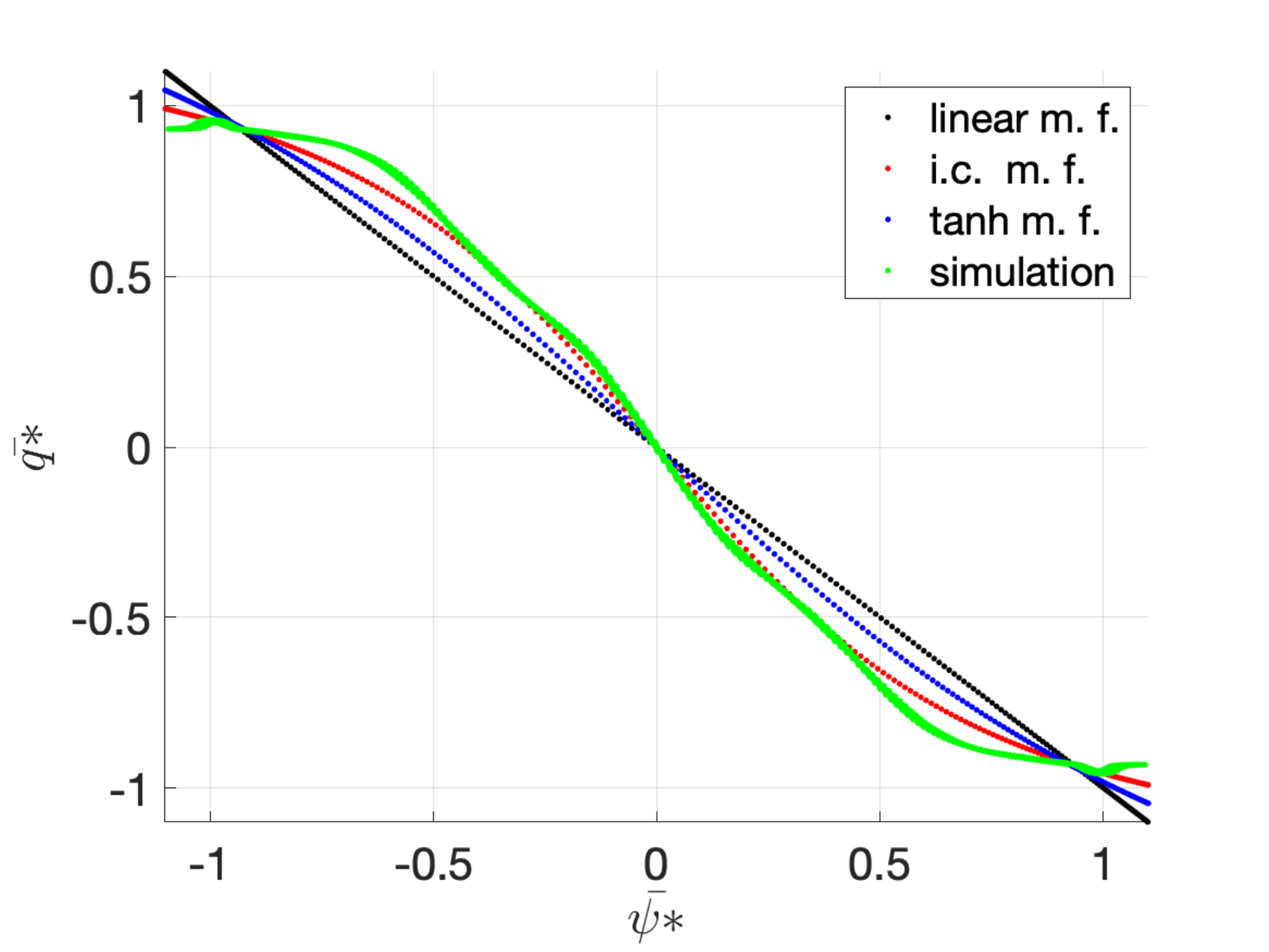}}}\\%
\caption{Left panels:  $q$ (colors) and $\psi$ (dashed line) at the equilibrium obtained with the adaped Turkington and Whitaker algorithm \cite{Turkington1996} and the mean field equation \eqref{tanhqp}. Right panels: functional relation between the same two fields for the case (a,b) $\alpha=1$, and (c,d) $\alpha=2$ obtained with different mean field relations.  The black line shows the linear mean field functional relation obtained using a Gaussian prior distribution into the Turkington and Whitaker algorithm, in blue the result obtained with the  tanh mean field equation, in red the mean field equation obtained solving numerically \eqref{mfe} with a  prior distribution equal to the initial global distribution as in Fig. \ref{fig:singpdf}, and finally in green line we report the result obtained with the simulation shown in Fig. \ref{fig:n8}}%
\label{fig:wt1}
\end{center}
\end{figure}
The equilibrium fields computed with the adapted Turkington and Whitaker algorithm by means of the mean field equation \eqref{tanhqp} are shown in  Fig. \ref{fig:wt1}a-c. Others equilibrium extrema have also been computed using the linear mean field equation, and solving numerically  the mean field equation \eqref{mfe} with a  prior distribution equal to the bimodal distribution obtained by the i.c.s under consideration (Fig. \ref{fig:singpdf}, black line). 
The features of the fields resemble the ones shown in \ref{fig:n8}, except for their orientation. The different orientation can however be explained by the invariance under rotations of the system. Nevertheless, significant  changes can be seen in the $\psi-q$ functional relation for different mean field equation (Fig. \ref{fig:wt1}b,d). In particular for $\alpha=1$, the $\psi-q$ relation shown in Fig. \ref{fig:n8} is in between the curves obtained using the linear mean field equation and \eqref{tanhqp}. This relation seems to be approximately linear and appears to differ from the curve obtained by considering a prior distribution equal to the true global initial vorticity distribution.
The opposite happens for the case $\alpha=2$ in which the simulation and the result obtained with the true prior distribution match well. In contrast to \cite{Venaille2015}, increasing $\alpha$ seems to bring the flow closer to the actual equilibrium state predicted by the theory.

In both cases the resulting $\psi-q$ relation is tanh-like and no transition to sinh-like relation is observed. However, as noted by \cite{Venaille2015}, increasing the value of $\alpha$ favors irreversible mixing of the global distribution of vorticity during the flow evolution with the emerging of little unmixed vortices.
These structures are related to sinh-like relations. In fact looking at the single snapshot of the final evolution of the field in particular for $\alpha=2$ in our simulation (not shown here), when $|\psi|>1$ the $\psi-q$ relation  deviates from the tanh-like shape in favour of the sinh-like relation.  Nevertheless, these little unmixed vortices should be seen as fluctuations over the equilibrium state and are removed by considering time averages. Increasing  $\alpha$ there are more  unmixed structures  and we can expect the statistical theory to fail.

\begin{figure}[hbt!]
\begin{center}
{\includegraphics[width=1\textwidth]{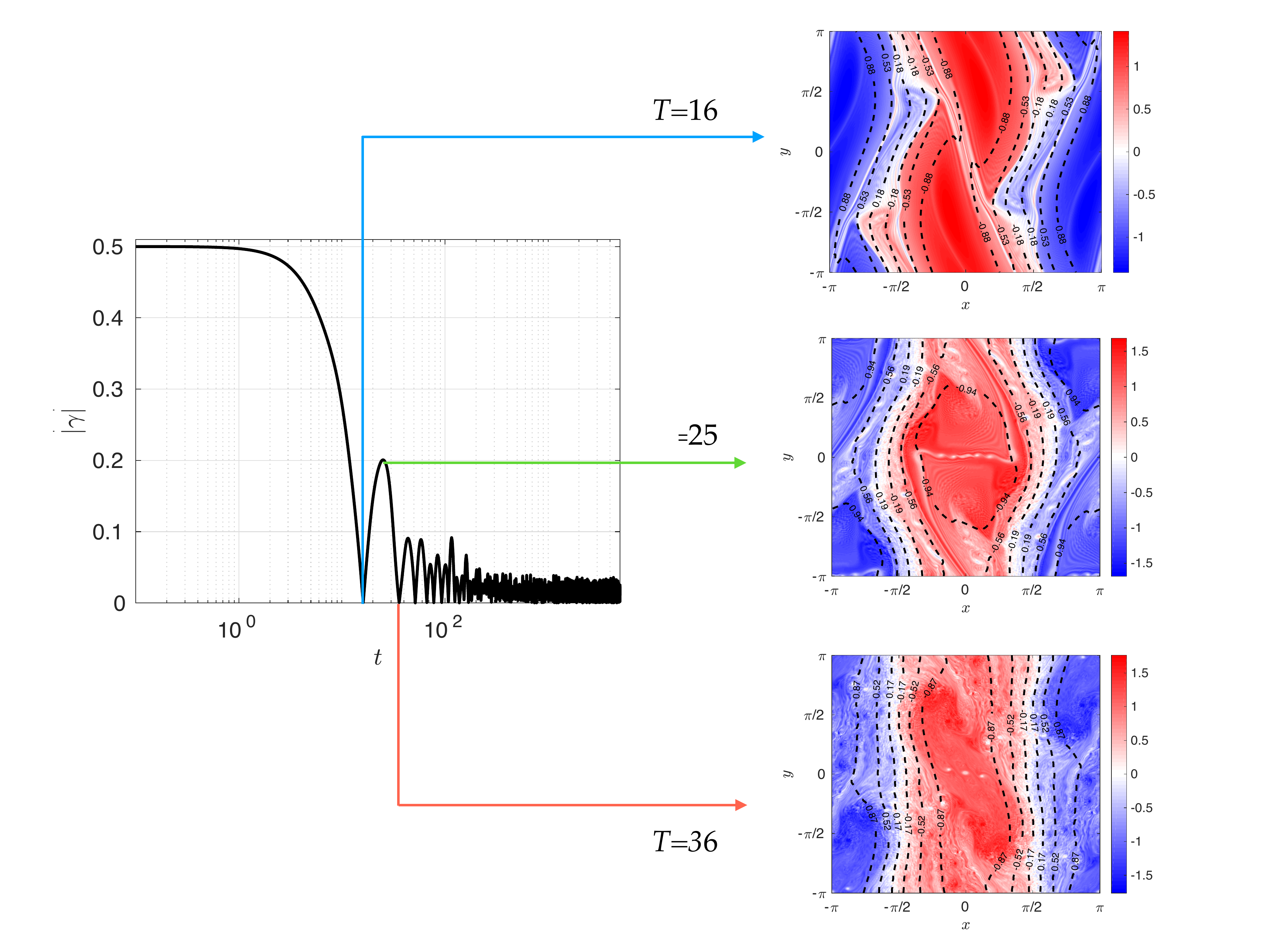}}
\caption{Behavior of the natural order parameter \eqref{gamma} in time. During the transition to the equilibrium the topology of the flow changes oscillating between a unidirectional flow and the formation of a vortex.}%
\label{fig:trans}
\end{center}
\end{figure}

In Fig. \ref{fig:trans} we investigate the transition to the equilibrium by means of the natural order parameter \eqref{gamma} for both $\alpha=1,2$. Around the time of the breaking of the filament we observe a monotonic decrease of $|\gamma|$. 
Consequently, the flow assumes a unidirectional structure and oscillations start to appear in the order parameter.
The first maxima in this oscillation is related to the formation of a vortex structure in the middle of the domain appearing at $T=25$,  followed by a unidirectional flow structure at $T=36$. These topological transitions are not sustained by  out of equilibrium dynamics (see e.g. \cite{Bouchet2009})  and they decay, leaving the system in a  state close to the equilibrium.
Small oscillations in $|\gamma|$ show that the stationary equilibrium is, however, never reached completely \cite{dritschel_1998}.
As in \cite{Venaille2015} our domain is double periodic with aspect ratio one. In this marginal case the actual extrema are degenerate and we can see a  competition between the extrema triggered by the chaotic motion around the broken filament.

\begin{figure}[hbt!]
\begin{center}
{\includegraphics[width=.8\textwidth]{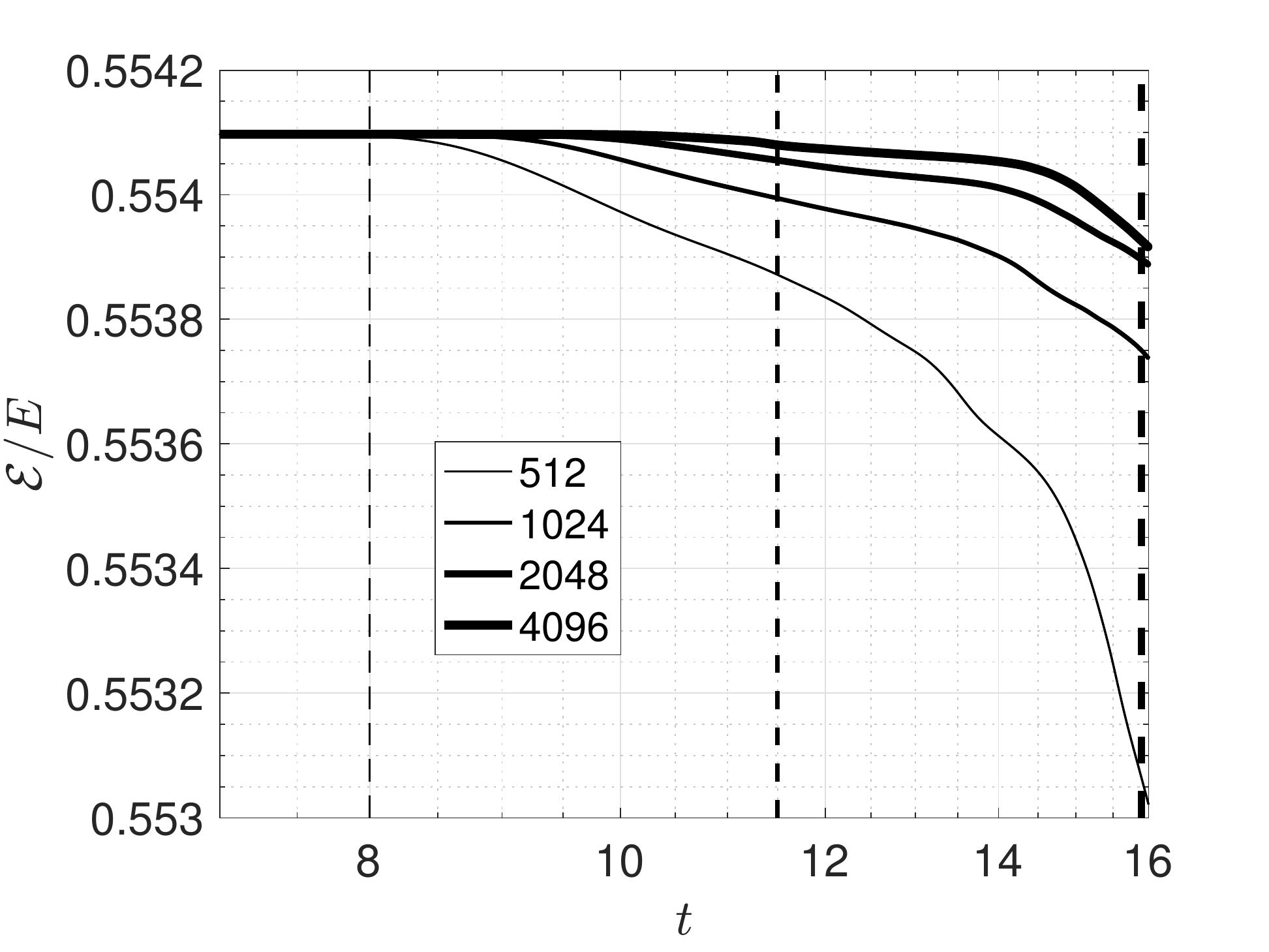}}%
\caption{Decay of $R$ for $\alpha=1$ and different resolutions. The vertical dashed lines represent the times $T=8$, $T=11.5$ and $T=15.9$ respectively.}%
\label{fig:res}
\end{center}
\end{figure}

\begin{figure}[hbt!]
\begin{center}
{\includegraphics[width=.8\textwidth]{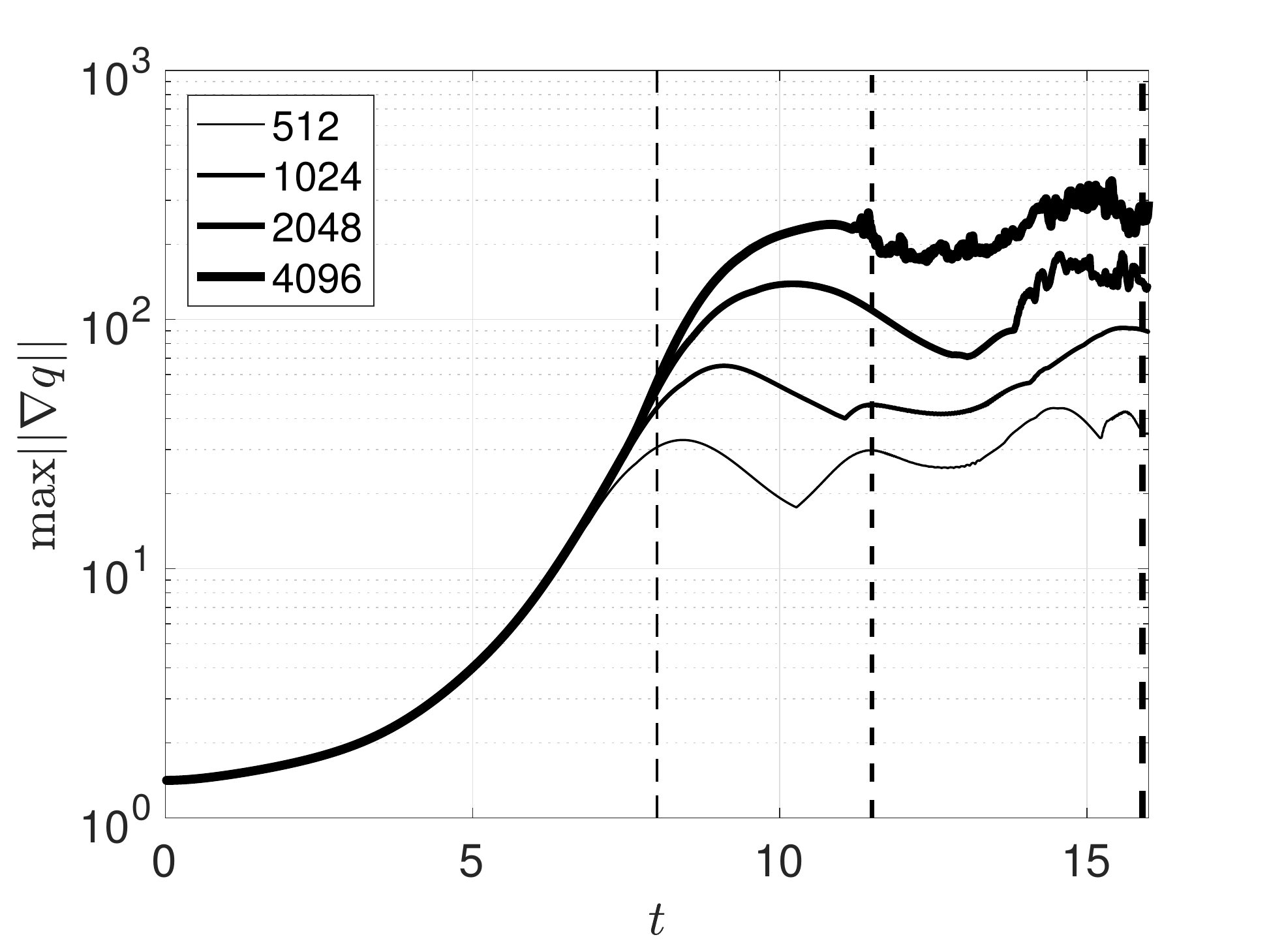}}%
\caption{Behavior of  $\max \lVert\nabla\theta\rVert$ for different resolutions. The vertical dashed lines represent the times $T=8$, $T=11.5$ and $T=15.9$ respectively.}%
\label{fig:maxgrad}
\end{center}
\end{figure}

These features do not appear in Fig. \ref{fig:n7} where the ratio $R$ decreases monotonically without the presence of oscillations.
For $\alpha=1$ and  $4096 \times 4096$ resolution, $R$ begins to  decay strongly
at $T \sim 11.5$, i.e. when the central filament breaks by secondary instabilities (Fig. \ref{fig:res}).
 It is interesting to compare the time evolution  of $R$ and of the quantity $\max \lVert\nabla\theta\rVert$, which enters the modified Beale-Kato-Majda criteria \eqref{eq:bkm}. 
Figure \ref{fig:maxgrad} shows that at $T \sim 11.5$  the $\max \lVert\nabla\theta\rVert$ starts to exhibit non-smooth oscillations associated with the breaking of the filament by secondary instabilities. 
These results show that \emph{the selective decay appears when the flow becomes unstable by secondary instabilities and these instabilities change its topology, allowing for the flow to evolve to a state with minimum values of $R$}.

The comparison of the results at $4096 \times 4096$ resolution with the integrations at lower resolution is in agreement with the observation that selective decay is scale-dependent. Lowering the resolution shows an anticipation of the selective decay, in agreement with the anticipated breaking of the central filament at lower resolution (Fig. \ref{fig:maxgrad}).


\subsection{\label{sec:level43} Perturbed Hyperbolic Saddle}
In order to explore the dependence of the results on the i.c.s, 
 we consider
an initial state of the form
\begin{equation}
q(\vec{z})=q_{hs}(\vec{z}) + \delta q_{rand}(\vec{z}),
\label{mixic}
\end{equation}
where $q_{hs}$ is the i.c.s \eqref{sic} normalized such that  max$(q_{hs})=1$, and $\delta$ is the amplitude of the perturbation. In the following simulations we assume $\delta=0,\, 0.1,\, 0.25,\, 0.5$.
The  random field $q_{rand}$   is obtained by the inversion of the stream function 
\begin{equation}
\hat{\psi}(\vec{k},0)\propto\exp[- (\left\lVert\vec{k}\right\rVert-k_0)^2]
\end{equation}
with $k_0=3$, after applying Gaussian noise,  and  is normalized so that max$(q_{rand})=1$, thus  $\delta$ measures the fraction of the maximum value of the perturbation in respect to the one of $q_{hs}$.
The simulations use  $512\times512$ points horizontal grid resolution.
The other parameters are the same as in the unperturbed simulations.
All the simulations are run until $T=10^{3}$.
The experiments are performed   only for $\alpha=1$.

\begin{figure}[hbt!]
\begin{center}
\subfloat[]{
{\includegraphics[width=.5\textwidth]{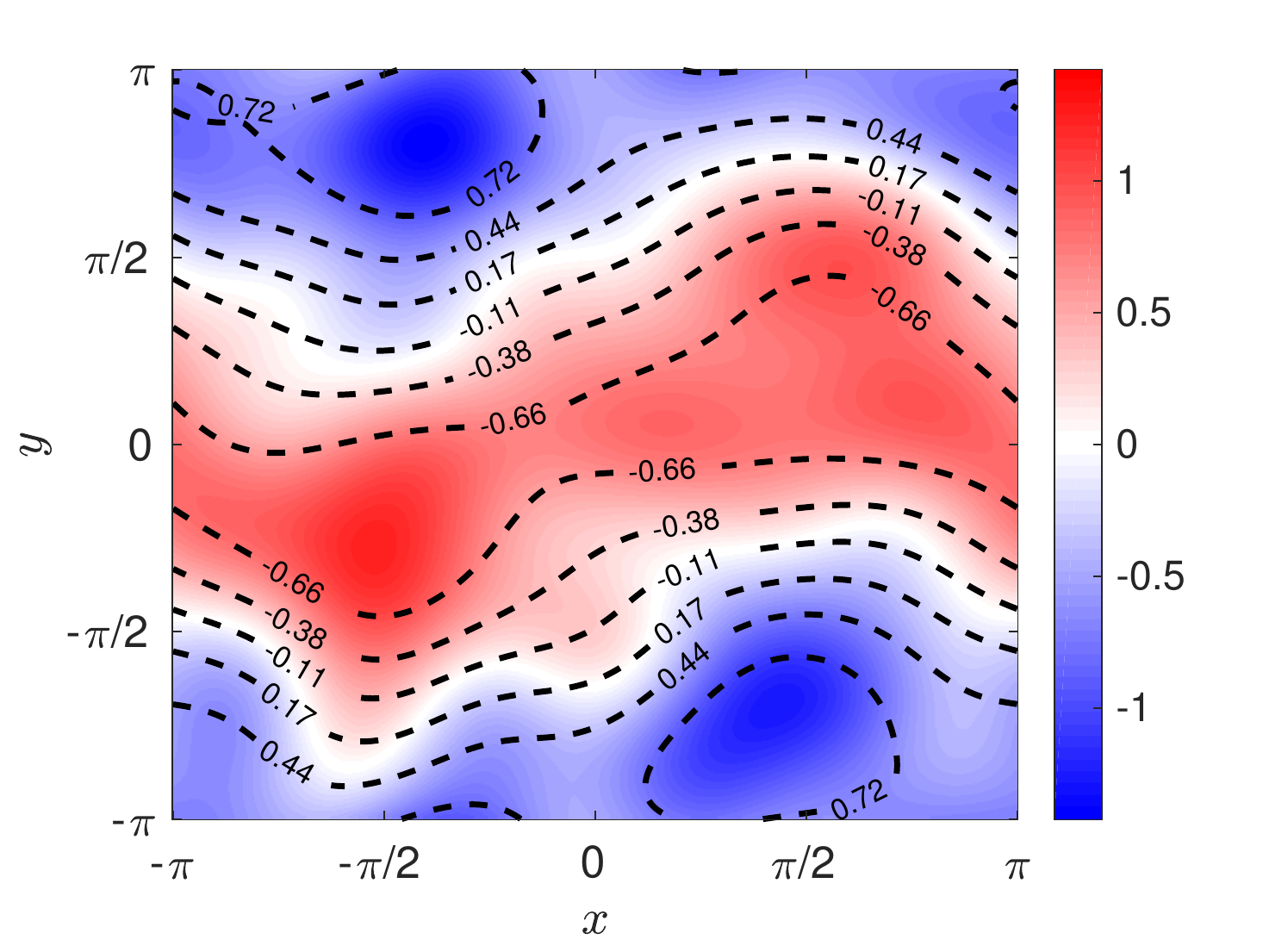}}}%
\subfloat[]{
{\includegraphics[width=.5\textwidth]{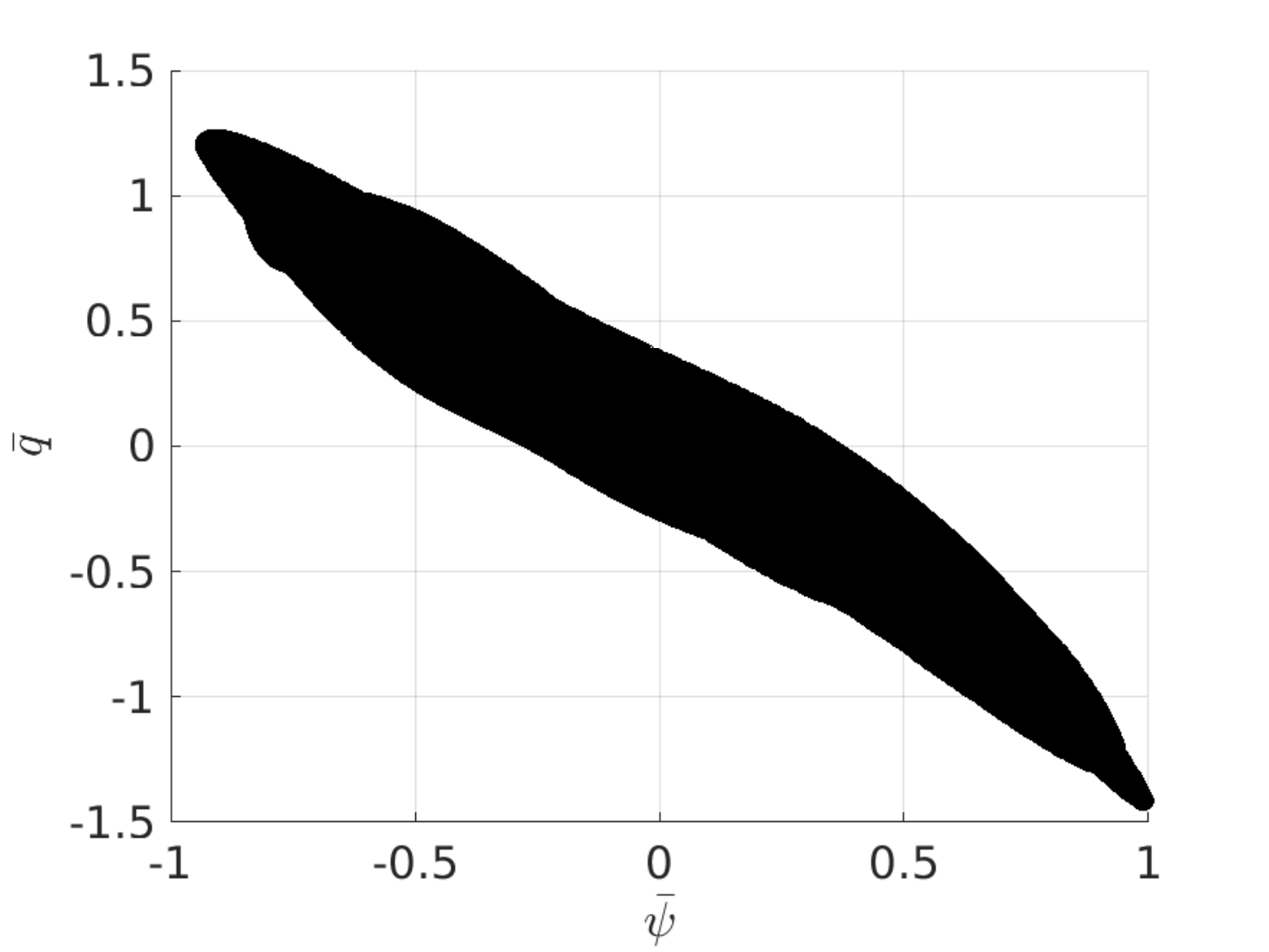}}}\\%
\subfloat[]{
{\includegraphics[width=.6\textwidth]{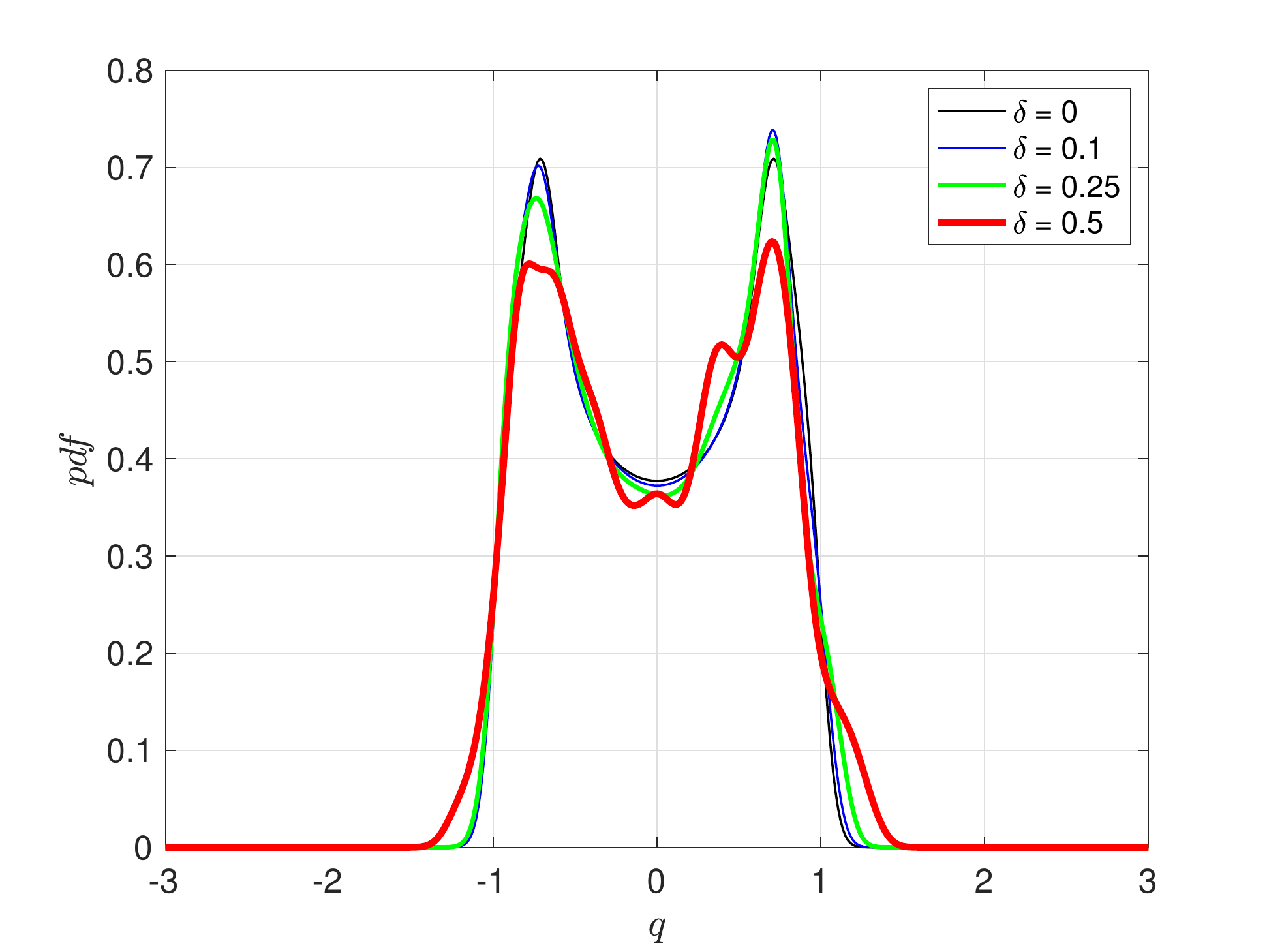}}}\\%
\caption{Initial conditions \eqref{mixic} with $\delta=0.5$. The grid resolution here is  $512\times512$ grid points. Panel (a) shows the $q$ field with the superimposed stream function (dashed line). Panel (b) shows the corresponding $\psi-q$ relation and panel (c) the initial global distribution of the $q$ field for every value of $\delta$. }%
\label{fig:mixic}
\end{center}
\end{figure}

Fig. \ref{fig:mixic} (a) shows \eqref{mixic} for $\delta=0.5$,  while panel (b) shows the functional relation $\psi-q$ for this i.c.
All the global initial pdfs are shown in panel (c) of the same figure. The perturbations introduce asymmetry  in the initial field.

\begin{figure}[hbt!]
\begin{center}
\subfloat[$\delta=0$]{
{\includegraphics[width=.5\textwidth]{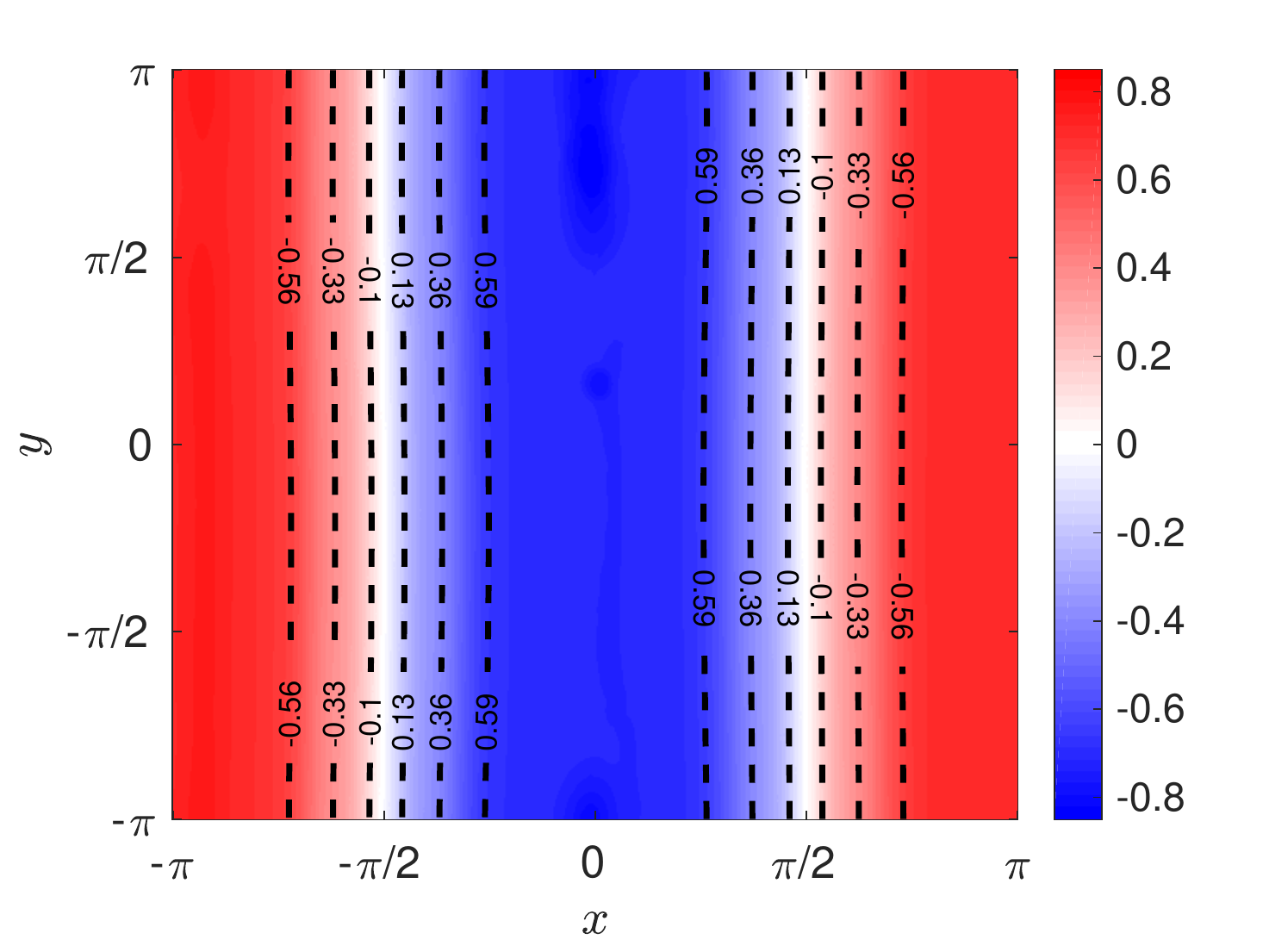}}}%
\subfloat[]{
{\includegraphics[width=.5\textwidth]{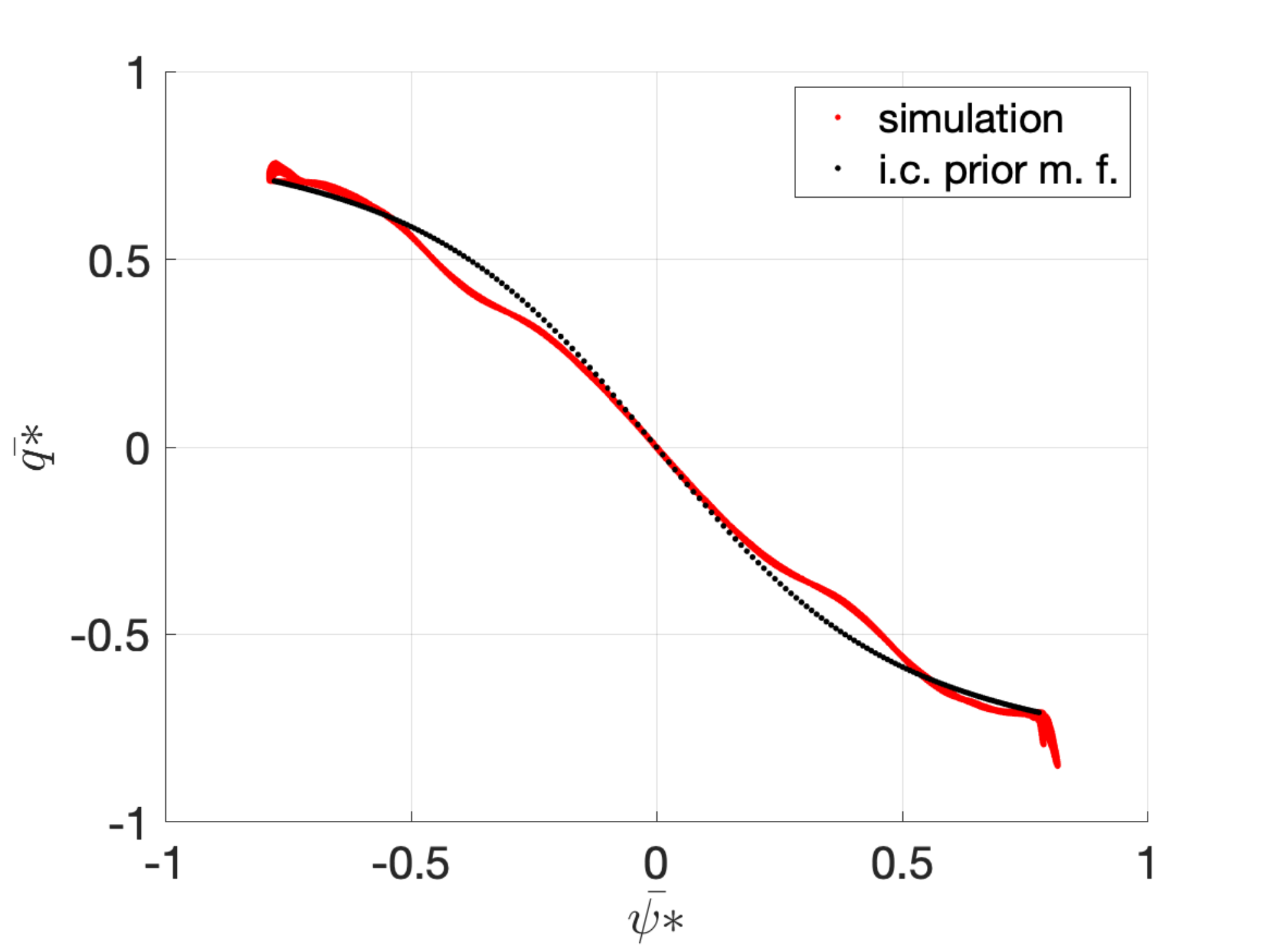}}}\\%
\subfloat[$\delta=0.5$]{
{\includegraphics[width=.5\textwidth]{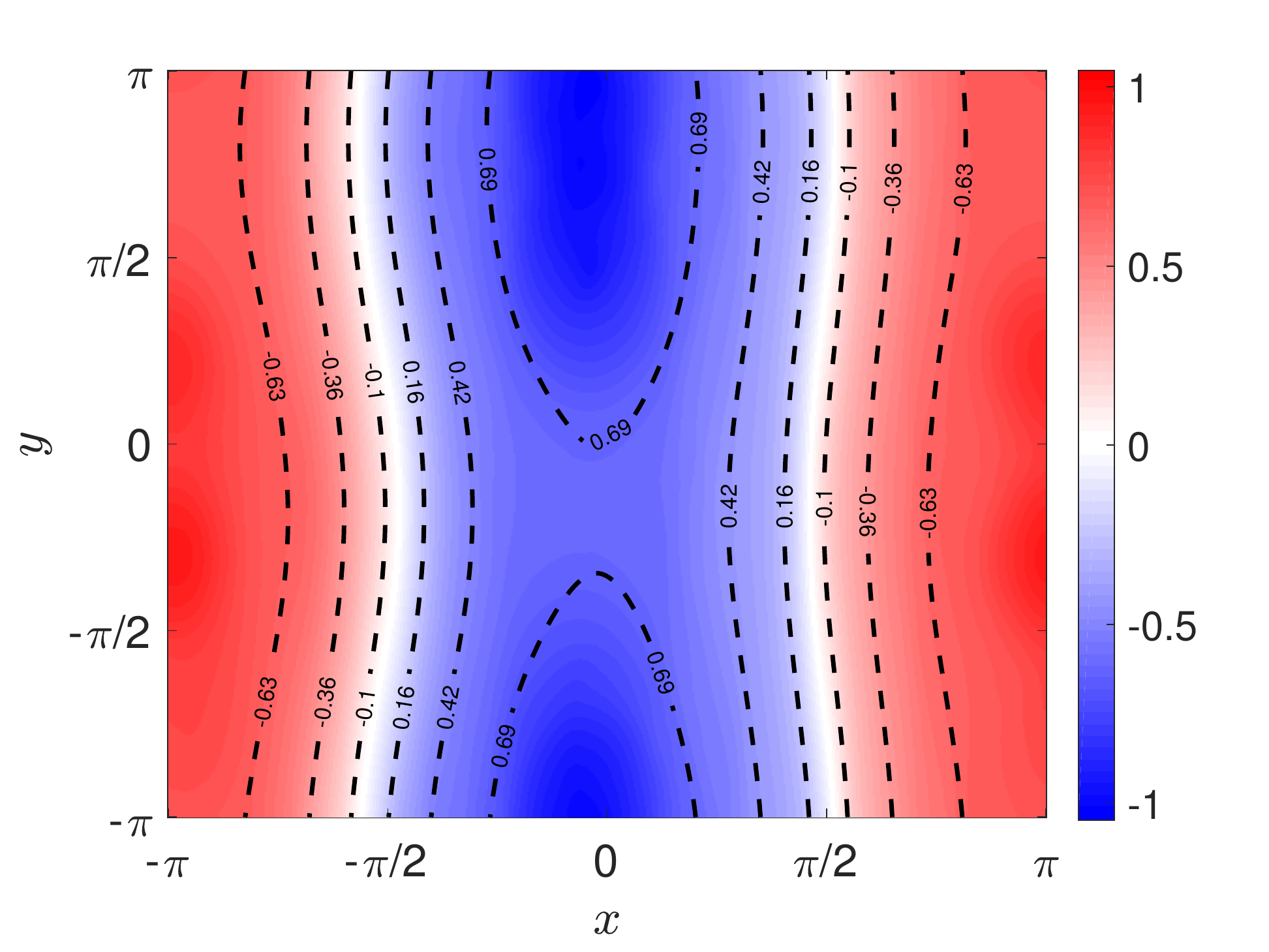}}}%
\subfloat[]{
{\includegraphics[width=.5\textwidth]{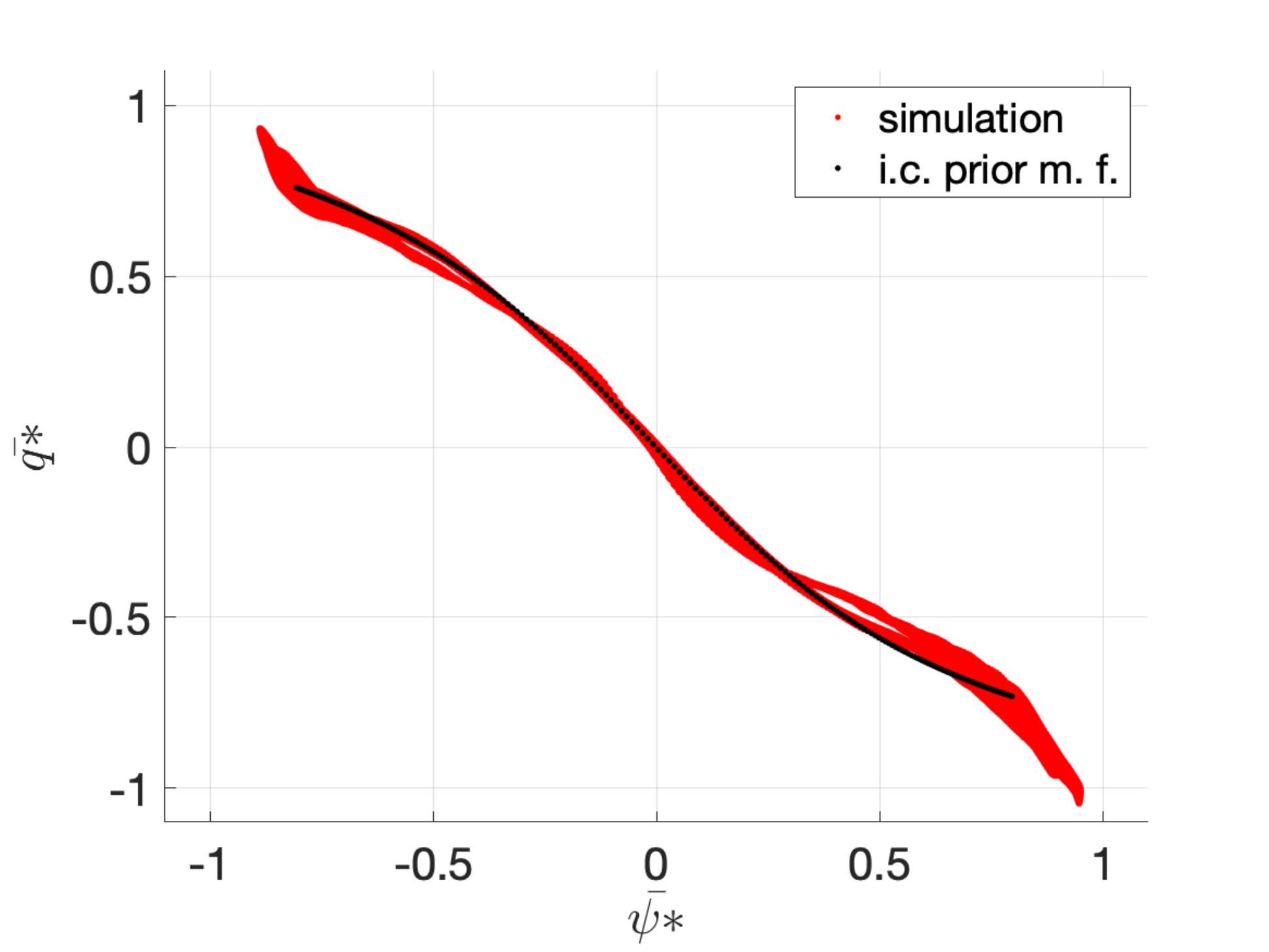}}}\\%
\caption{Left panels: time average of  $q$ and $\psi$ fields for $\delta=0$, panel (a) and $\delta=0.5$, panel (c). Right panels: functional relation between the same two fields, red lines, and the functional relations suggested by the theory and computed by means the iterative algorithm, black lines.  For this long run  the grid resolution used is set to  $512\times512$ grid points.}%
\label{fig:mixend}
\end{center}
\end{figure}

Fig. \ref{fig:mixend} (a, c) shows the  averaged final fields for the case $\delta=0$ and $\delta=0.5$. All the experiments exhibit a final unidirectional flow as equilibrium state. The evident difference is the ellipsoidal structure in  the central part of the domain for $\delta=0.5$. The perturbation tends to favor the formation of vortices that result to be more persistent increasing the $\delta$ value. The persistent vortex moves in the meridional direction  and a cleaner unidirectional flow can be seen by taking a longer time average. Figs. \ref{fig:mixend}b,d show the functional $\psi-q$ relation for the averaged field of the two simulations considered and the ones obtained by means of the initial global vorticity probability density as prior distribution in the Turkington and Whitaker algorithm. 
The $\psi-q$ relation for the $\delta=0.5$ case shows a small double branch in the simulation, indicating that the equilibrium has not being really  reached, and that the presence of persistent vortices can be removed by considering a longer time average. It should be noted that the fields are relaxing on the state predicted by the theory.

\begin{figure}[hbt!]
\begin{center}
\subfloat[]{
{\includegraphics[width=.5\textwidth]{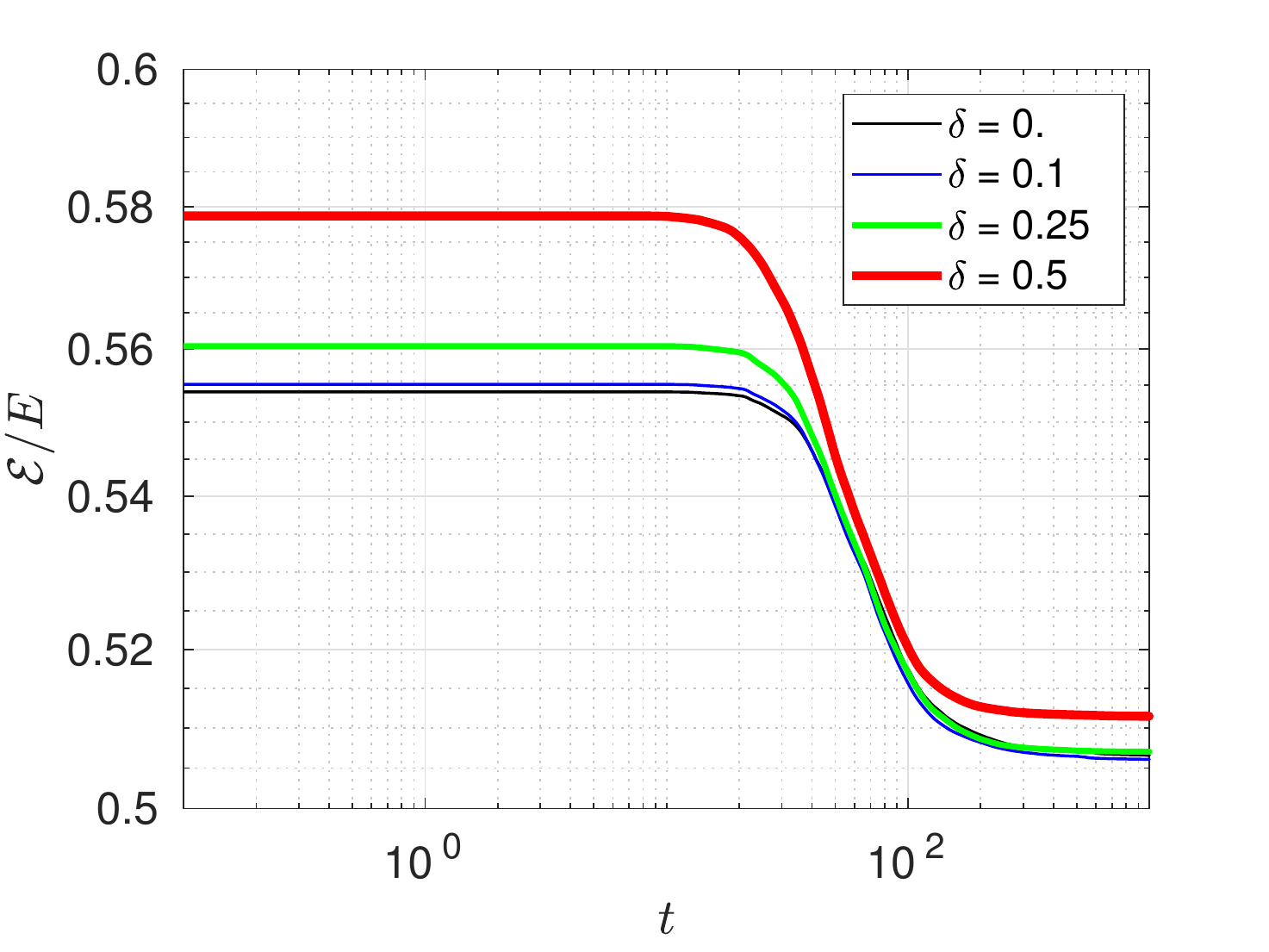}}}%
\subfloat[]{
{\includegraphics[width=.5\textwidth]{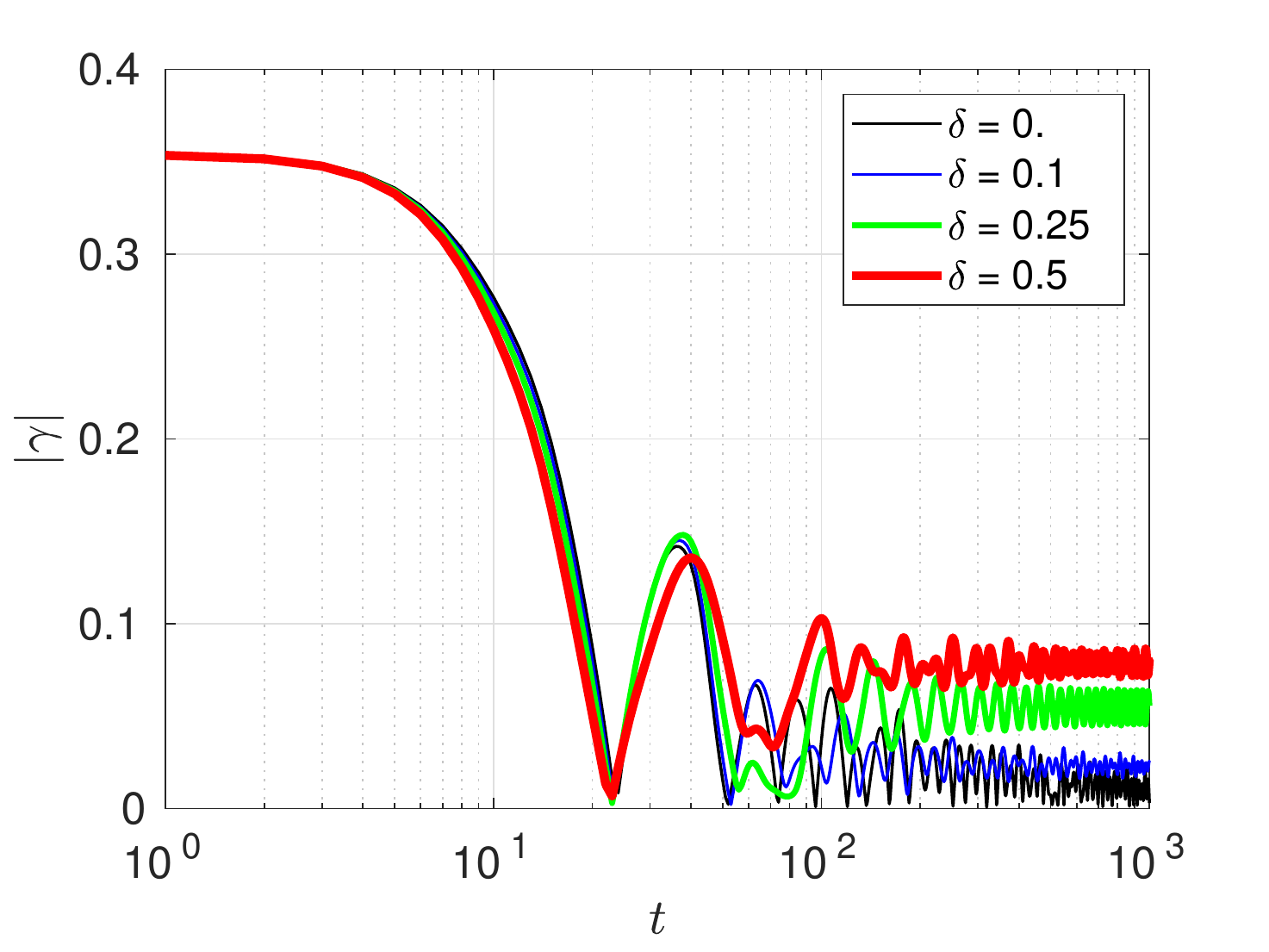}}}\\%
\subfloat[]{
{\includegraphics[width=.6\textwidth]{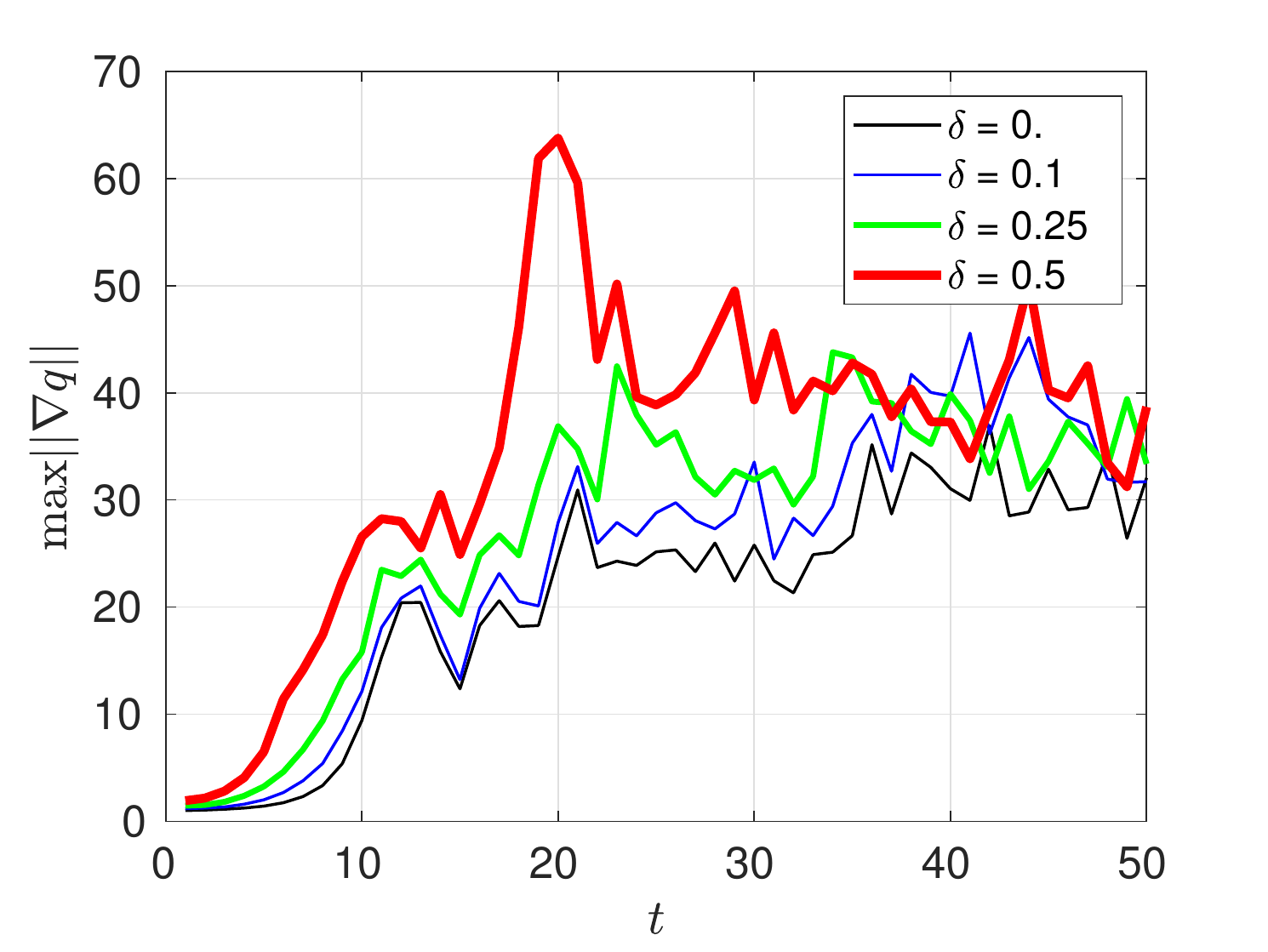}}}\\%
\caption{Comparison, for different values of $\delta$, of $R$, $|\gamma|$ and max$||\nabla q||$ respectively in panel (a), (b) and (c).}%
\label{fig:rggmix}
\end{center}
\end{figure}

The ratio $R$ decays monotonically for all  the cases (Fig. \ref{fig:rggmix}a). All the experiments start with a different value of $R$, however, at the end of the simulation only the case with $\delta=0.5$  seems to remain well separated from the others. The emergence of more persistent vortices is highlighted by the order parameters (Fig. \ref{fig:rggmix}b). 
The time evolution of $|\gamma|$ behaves almost exactly in the same way for all the experiments until it reaches  the zero value for the first time.
 This can be explained by considering that an hyperbolic system is not strongly influenced by a small bounded random perturbation \cite{Ano67,Bowen1975}. However, after the breaking of the filament, the flow loses its hyperbolicity and it evolves to different final states. 
The final fields predicted by the theory have the same rotated unidirectional features (not shown here) as in the previous section.
Larger values of $\delta$ seem to favor the formation of vortices, and thus a sinh-like $\psi-q$ relation. The order parameter seems to emphasize the  differences in the final state better. 
Fig.  \ref{fig:rggmix} (c) shows the behavior of the maximum of the gradient of $q$ for the different simulations. 
Increasing the amplitude of the perturbation acts to halt  the growth of the gradient and allows for a faster decay of the order parameter.


\section{\label{conc}Conclusions}

In this work we have investigated a family of turbulent models described by the generalized Euler equations studying their long time statistical equilibrium, and their relaxation to the equilibrium. 
Similarly to the  MRS theory \cite{Miller1990,robert_sommeria_1991} we invoked the maximization of appropriate entropy functionals translating the research of the probability measure  in an optimization problem with constraints given by opportunely chosen conserved quantities of the system.  This allowed the definition of  the mean field equations, that play the role of constraints for some possible final  state for the turbulent flow. It is important to stress that the set of stable states could be much larger than that the one arising by the equilibrium statistical theories, implying thus the possible existence of other invariant measures. Other problems in the application of this statistical theory arise also in regard to the ergodic hypothesis and the conservation of some of the quantities used as constraints for other geometries \cite{Dritschel2015-1,Dritschel2015-2}.
The theory suggests, under particular conditions, a selective decay principle for all the family of generalized Euler equations.
Despite the fact that the result is related to  a Gaussian prior distribution,  this principle seems to be more general as highlighted also by the numerical simulations.
We have investigated the role of the selective decay in the transition to the final state for i.c.s given by a hyperbolic saddle for the case of $\alpha=1$, which is a possible candidate for the emergence of singularities in SQG. In this case, the violent decay of the ratio between the generalized enstrophy and energy corresponds to the rise of secondary instabilities and the subsequent breaking of the central filament of the flow, suggesting thus a relationship between the selective decay and the change of topology of the flow.   

The transition to the equilibrium has been explored by means the natural order parameter introduced by \cite{Bouchet2009,Bouchette2012}. Since the domain for the numerical simulations is double periodic with aspect ratio one, the extrema for the statistical theory are degenerate. The use of the natural order parameter allows to visualize the jumps between these states in the flow topology induced by the filament breaking.

The equilibrium states are compared with the ones obtained by the theory and computed with an adapted iterative algorithm \cite{Turkington1996}. Although the final fields denote similar features, the functional relation $\psi-q$ denotes a difference between $\alpha=1$ and $\alpha=2$. In particular  the nonlinearity in the $\psi-q$ relation is more accentuated incrementing the value of $\alpha$. The comparison with the theory also shows that SQG does not reach the steady states predicted by the equilibrium, although the final unidirectional vorticity fields are similar. 

Contrarily to \cite{Venaille2015}, starting from the hyperbolic saddle i.c.s, but using the same double periodic domain with aspect ration one, we can not see a  transition in the shape of the functional relation $\psi-q$ from a tanh-like to a sinh-like shape. Increasing $\alpha$, it is possible to observe the emergence of persistent vortices that tend to deform the  $\psi-q$ relation. 
These vortices are however filtered by the time average used to determine the equilibrium state in the simulations. Furthermore, their effects on the mean fields are small.

Simulations starting from perturbed  hyperbolic saddle i.c.s for $\alpha=1$ show that the natural order parameter is not really sensitive to the perturbation until the breaking of the filament. This is because the i.c.s are hyperbolic while  after the  formation of vortices the system can evolve in different ways. In particular, increasing the perturbation amplitude the system is not able to decay to one of the extrema, but because of  the presence of more persistent vortices it results to be in a combination of different states. In all these cases the selective decay is always present denoting the importance of this principle.

Others physical systems, in which the potential vorticity and the stream function are related by means of different operators than the one discussed here, can be explored with the same techniques used in this work.
An example is given by  the Surface Semi Geostrophic (SSG) model \cite{SSG,ragone2016study}, where the operator connecting the potential vorticity and the stream function is a nonlinear Monge-Ampere operator that can be used to describe ageostrophic motion in geophysical flows. The role of the selective decay in shaping these dynamics is thus important for the understanding of the turbulence of the climate system.

\begin{acknowledgements}
GC and GB  warmly thank the referees whose comments greatly improved the manuscript. This work was partially funded by the research grant DFG BA5068/9-1.
\end{acknowledgements}

\clearpage
\appendix
\section{\label{sec:level2} Generalized Point-Vortex Statistical Theory for Local Dynamics} 

In the point vortex approximation \cite{Onsager1949} the vorticity field is substituted by a set of localized point vortices, which is the analogous of replacing a continuous mass distribution by a set of localized material points.
The stability of SQG  vortices  has been studied by e.g. \cite{carton2009instability,dritschel2011exact,harvey2011perturbed,harvey2011instability,bembenek2015realizing,carton2016vortex,Badinpoulin2018}, while SQG point vortices have been studied by e.g. \cite{BadinBarry2018,lim2001point,taylor2016dynamics}.
For the statistical dynamics of point vortices in 2D fluid dynamics see e.g. \cite{Onsager1949,Weiss1998,Chavanis1999,buhler2002statistical,Chavanis2009,esler2017equilibrium}. 

The velocity statistic for point vortices of the generalized Euler equation has been studied in \cite{Conti2019}.
In particular, \cite{Conti2019} showed that while for $\alpha=2$ the thermodynamic limit for the point vortex system does not exist due to logarithmic divergence with the number of vortices \cite{jimenez_1996,Weiss1998}, this limit is instead defined for $\alpha \neq 2$. 

Recently some renewed interest  has been shown on systems of point vortices for the generalized Euler dynamics \cite{Geldhauser2018PointVF}. In particular, for $\alpha=1$ a statistical measure with the only use of the energy constraint has been found with the help of regularized functions. In the following, we show that if we can consider another constraints,  the angular impulse, this regularization is not necessary.

In the point-vortex approximation we consider an active scalar concentrated in some points, so that
it can be written in terms of the sum of Dirac delta functions, i.e a point-vortex located in the position
$\vec{z}_0=(x_0,y_0)$ generates an active scalar field of the form
\begin{equation}
q(\vec{z})=\gamma \delta(\vec{z}- \vec{z}_0)
\label{pvdef}
\end{equation}
where  $\gamma$ here is the generalized circulation.
 Solving \eqref{invp} it is possible to find the stream function and then,
by means of the usual derivation, the velocity field in the domain considered.
The stream function is easily written in terms of the Green's function of the problem considered
\begin{equation}
\psi(\vec{z})=\int G_{\alpha}(\vec{z},\vec{z}')q(\vec{z}')\,\mathrm{d}\vec{z}',
\end{equation} 
where
\begin{subequations}
\begin{align}
G_{2} \left( \vec{z},\vec{z}' \right)\,&\,=\,\frac{1}{2\pi}\ln \left(\left\lVert\vec{z}-\vec{z}'\right\rVert \right) \,, \,\,\,\qquad \alpha=2\,,\label{g1}\\
G_{\alpha} \left( \vec{z},\vec{z}' \right)\,&\,=\,\mathrm{\phi}(\alpha)\left\lVert \vec{z}-\vec{z}' \right\rVert^{\alpha-2}\,, \qquad \alpha\neq2\,,\label{g2}
\end{align}
\end{subequations}
and
\begin{equation}
\mathrm{\phi}(\alpha)=-\left\{2^\alpha\left[\mathrm{\Gamma}\left(\frac{\alpha}{2}\right)\right]^2 \sin\left(\frac{\alpha\pi}{2}\right)\right\}^{-1}\,,
\end{equation}
see for example \cite{BadinBarry2018,Iwayama2010}.

When $\alpha>3$ the effect of one vortex on another increases with distance, and hence this case is considered unphysical \cite{BadinBarry2018}. Only the interval $\alpha \in (0,\,3]$ will thus be considered.

\subsection{\label{sec:level2_1} Generalized Point-Vortex Equations} 

For the point-vortex approximation the resulting dynamics is Hamiltonian and is characterized by the conservation of energy, as well as the linear and angular momentum \cite{helmholtz1858}.  See e.g. \cite{aref2007point,marchioro2012mathematical,newton2013n} for reviews and \cite{chapman1978ideal,badin2018variational} for a discussion on symmetries and conservation laws. 

Given $N$ point vortices with circulations $\gamma_i$, $i=1,\dots,N$, and with 
 the coordinates of the vortices locations given by 
$\bm{z}=(x_1,y_1,\dots,x_N,y_N)=(\vec{z}_1,\dots,\vec{z}_N)$,
the equations of motion for the  vortices can be written as
\begin{equation}
\dot{\bm{z}}=\bm{{\mathrm D}}_{\alpha}^{-1}\bm{{\mathrm J}}\nabla_{\bm{z}}H_{\alpha}.
\label{eqHcompact}
\end{equation}
where $\bm{{\mathrm D}}_{\alpha}=\diag(\gamma_1,\gamma_1\dots,\gamma_N,\gamma_N)$ when $\alpha=2$ and $\bm{{\mathrm D}}_{\alpha}=\diag(\gamma_1/(2-\alpha),\gamma_1/(2-\alpha)\dots,\gamma_N/(2-\alpha),\gamma_N/(2-\alpha))$ when $\alpha\neq2$, and $\bm{{\mathrm J}}$ is a block diagonal matrix with blocks given by symplectic matrices \cite{badin2018variational}, and   
\begin{subequations}
\begin{align}
H_{2} &= -\frac{1}{4\pi}\sum_{\substack{i,j=1\\ j\neq i}}^N\gamma_i\gamma_j\log\left\lVert \vec{z}_i - \vec{z}_j\right\rVert, \\
H_{\alpha} &= -\frac{\mathrm{\phi}(\alpha)}{2}\sum_{\substack{i,j=1\\ j\neq i}}^N\frac{\gamma_i\gamma_j}{\left\lVert \vec{z}_i - \vec{z}_j\right\rVert^{2-\alpha}}\,, \qquad \alpha\neq2\,,
\end{align}
\label{Ha}
\end{subequations}
is the Hamiltonian function, that is a  conserved quantity representing the energy of the system.

It is possible to proof the conservation of other quantities apart from the Hamiltonian one \cite{BadinBarry2018}. Among the others, the angular impulse defined as
\begin{equation}
I=\sum_{\substack{i=1}}^N \gamma_i (x_i^2 +y_i^2)=\sum_{\substack{i=1}}^N \gamma_i \left\lVert \vec{z}_i \right\rVert^{2},
\label{I}
\end{equation}
 plays a special role in defining a statistical measure. Other invariants are the linear impulse or the squared relative angular momentum with respect to the center of circulation that is a combination of the linear invariants and $I$. However, it is possible to show that these constants of motions do not add other information to the statistical measure. The linear invariant, in fact, can be neglected with a suitable change of variable. Note that only the Hamiltonian includes the information about the locality of the interaction $\alpha$.


\subsection{\label{sec:level2_2} Statistical Formulation}
Since the system considered is Hamiltonian, it possesses the Liouville property. Furthermore, since the systems considered exhibits chaos due to their nonlinearity, we will assume ergodicity.  In order to find a  non-trivial probability density function $p({\bm z})$ involving the maximum entropy principle  we need to use some constraints, in particular the Hamiltonian, that incorporate the information on the strength of the interaction $\alpha$, and the angular impulse $I$. 
In the following we also assume, without loss of generality, that $\gamma_k=1$ for $1\le k\le N$.

Considering an entropy functional defined as
\begin{equation}
\mathcal{S}(p)=-\int_{\mathcal{R}^{2N}} p({\bm z})\log p({\bm z})\,\rm{d}{\bm z},
\label{entropy}
\end{equation}
where the Lebesgue measure is considered, we need to find $p^*$ so that
 \begin{equation}
 \mathcal{S}(p^*)=\max_{p\in\mathcal{C}} \mathcal{S}(p).
 \end{equation}
 As explained above the constraints used are
\begin{equation}
\mathcal{C}=\mathcal{C}_0\cap\mathcal{C}_{\alpha}\cap\mathcal{C}_I
\end{equation}
where
\begin{subequations}
\begin{align}
&\mathcal{C}_0=\{p({\bm z}) \,\lvert\, p({\bm z})\ge0 \wedge \int_{\mathcal{R}^{2N}} p({\bm z})\,\rm{d}{\bm z}=1\}\\
&\mathcal{C}_{\alpha}=\{p({\bm z}) \,\lvert\,  \int_{\mathcal{R}^{2N}} H_{\alpha}({\bm z})p({\bm z})\,\rm{d}{\bm z}=\bar{H}_{\alpha}\}\\
&\mathcal{C}_{I}=\{p({\bm z}) \,\lvert\,  \int_{\mathcal{R}^{2N}} I({\bm z})p({\bm z})\,\rm{d}{\bm z}=\bar{I}\}.
\end{align}
\end{subequations} 
If  $\lambda_\alpha,\,\lambda_I$ are the Lagrange multipliers for the constraints related to the conserved quantities $H_{\alpha}$ and $I$ the solution derived is then
 analogous to the Gibbs measure in statistical mechanics with 
\begin{equation}
p^*({\bm z})=\mathcal{G}_{{\bm \lambda},N}({\bm z})=Z_{N}^{-1}\exp\left( -\lambda_\alpha H_{\alpha}({\bm z})-\lambda_I I({\bm z})\right), 
\label{Gibbs}
\end{equation}
with 
\begin{equation}
\begin{split}
Z_{N}&=\int _{\mathcal{R}^{2N}} \exp \left( {\frac{\lambda_\alpha\phi(\alpha)}{2} \sum_{\substack{i,j=1\\ j\neq i}}^N\frac{1}{\left\lVert \vec{z}_i-\vec{z}_j \right\rVert^{2-\alpha}}}- \lambda_I\sum_{\substack{i=1}}^N  \left\lVert \vec{z}_i \right\rVert^2 \right)\,\rm{d}{\bm z},
\end{split}
\end{equation}
and in order to satisfy the integrability $\lambda_\alpha \phi(\alpha)<0$, and $\,\lambda_I>0$. 

It is now clear that the use of the linear momentum in the constraints would result in linear terms in the measure that could be eliminated by using a suitable coordinates system.
%
%
%


The derivation of a mean field equation for the stream function for the general $\alpha$-model is identical to the one for the case $\alpha=2$ \cite{caglioti1992,caglioti1995,majda_wang_2006}.  Replacing $\lambda_\alpha$ by $\lambda_\alpha/N$, integrating \eqref{Gibbs} over $N-1$ variables to find the marginal distribution of a single vortex  $p^*(\vec{z})$ with $\vec{z}\in\mathcal{R}^2$, and  taking the thermodynamic limit  $N\to\infty$ \cite{Conti2019} we get the following \emph{mean field equation}
\begin{equation}
-(-\Delta)^{\alpha/2}\psi^*=p^*(\vec{z})=\frac{e^{-\lambda_\alpha \psi^*(\vec{z}) -\lambda_I \left\lVert\vec{z}\right\rVert^2} }{\int _{\mathcal{R}^{2}} e^{-\lambda_\alpha \psi^*(\vec{z}) -\lambda_I \left\lVert\vec{z}\right\rVert^2}{\rm{d}}\vec{z}}\,.
\label{mfpv}
\end{equation}

Note that if we consider the following partition function
\begin{equation}
\mathcal{Z}_\alpha=\int _{\mathcal{R}^{2}} e^{-\lambda_\alpha \psi(\vec{z}) -\lambda_I \left\lVert\vec{z}\right\rVert^2}{\rm{d}}\vec{z}\,,
\end{equation}
the $n$-momentum of the stream function is derived as
\begin{equation}
\langle \psi(\vec{z})^n)\rangle_{\mathcal{G}_{\bm{\lambda}}} = \frac{(-1)^n}{\mathcal{Z}} \frac{ {\rm{\partial}}^n} { {{\rm{\partial}} }\lambda_\alpha^n} \mathcal{Z}_\alpha\,.
\end{equation}

With a formally identical derivation as for $\alpha=2$, it is possible to obtain a mean field equation for the generalized $\alpha$-model of turbulence. However, if the angular impulse constraint could be neglected for $\alpha\ge2$, that is for $2$D turbulence and for non-local turbulence model, for the case of local turbulence $\alpha<2$ is fundamental to reach the convergence of the probability measure. This is related to a fundamental change in the topology of the system hidden in the Green's function of the problem and then in the corresponding stream function.

Note that for different geometries, for example on the sphere,  the ergodic hypothesis and the capability of the system of reaching a steady flow might be not valid, see e.g. \cite{Dritschel2015-1,Dritschel2015-2}.

\bibliographystyle{spphys}       


\end{document}